\documentclass[12pt, onecolumn, draftclsnofoot]{IEEEtran}
\usepackage{cite}
\usepackage{graphicx} 
\usepackage{epsf}
\usepackage{subfigure}
\usepackage{amsmath}
\usepackage{comment}
\usepackage{amssymb}
\usepackage{array}
\usepackage{setspace}
\usepackage{amsthm,amssymb}
\usepackage[ruled,lined]{algorithm2e}
\usepackage{multirow} 
\usepackage{tabularx}
\usepackage{longtable}
\usepackage{xfrac}
\usepackage{breqn}
\usepackage{bbm}
\usepackage{enumerate}
\usepackage{algorithmic}
\usepackage{dsfont}
\usepackage{pifont}
\usepackage{float}
\usepackage{diagbox}
\usepackage{mathtools}
\usepackage{adjustbox}
\usepackage{color}
\usepackage{ctable}
\usepackage{textcomp}

\graphicspath{{Fig/},{fig/}}

\newcommand{\tabincell}[2]{\begin{tabular}{@{}#1@{}}#2\end{tabular}}
\newcommand{\cmark}{\ding{52}}%

\begin{document}

% paper title
\title{ Five Facets of 6G: Research Challenges and Opportunities}

\author{Li-Hsiang Shen, Kai-Ten Feng, and Lajos Hanzo$\dagger$ \\
Department of Electronics and Electrical Engineering \\
National Yang Ming Chiao Tung University, Hsinchu, Taiwan\\
$\dagger$Department of Electronics and Computer Science \\ University of Southampton, UK\\
gp3xu4vu6.cm04g@nctu.edu.tw, ktfeng@nycu.edu.tw, and lh@ecs.soton.ac.uk}

%\author{\IEEEauthorblockN{Li-Hsiang Shen, Kai-Ten Feng, and Lajos Hanzo}\\
%\IEEEauthorblockA{Department of Electronics and Electrical Engineering \\
%National Yang Ming Chiao Tung University, Hsinchu, Taiwan\\
%$\dagger$ Department of Electronics and Computer Science \\ University of Southampton\\
%
%gp3xu4vu6.cm04g@nctu.edu.tw, ktfeng@nycu.edu.tw, and lh@ecs.soton.ac.uk}}

%\author{\authorblockN{}\authorblockA{Department of Electrical Engineering\\ National Chiao Tung University, Hsinchu, Taiwan\\
%				\IEEEauthorrefmark{1}Department of Electronic Engineering\\ National Taipei University of Technology, Taipei, Taiwan\\}}

\maketitle

\begin{abstract}
Whilst the fifth-generation (5G) systems are being rolled out across the globe, researchers have turned their attention to the exploration of radical next-generation solutions. At this early evolutionary stage we survey five main research facets of this field, namely {\em Facet~1: next-generation architectures, spectrum and services, Facet~2: next-generation networking, Facet~3: Internet of Things (IoT), Facet~4: wireless positioning and sensing, as well as Facet~5: applications of deep learning in 6G networks.} In this paper, we have provided a critical appraisal of the literature of promising techniques ranging from the associated architectures, networking, applications as well as designs. We have portrayed a plethora of heterogeneous architectures relying on cooperative hybrid networks supported by diverse access and transmission mechanisms. The vulnerabilities of these techniques are also addressed and carefully considered for highlighting the most of promising future research directions. Additionally, we have listed a rich suite of learning-driven optimization techniques. We conclude by observing the evolutionary paradigm-shift that has taken place from pure single-component bandwidth-efficiency, power-efficiency or delay-optimization towards multi-component designs, as exemplified by the twin-component ultra-reliable low-latency mode of the 5G system. We advocate a further evolutionary step towards multi-component Pareto optimization, which requires the exploration of the entire Pareto front of all optiomal solutions, where none of the components of the objective function may be improved without degrading at least one of the other components.

\end{abstract}


\section{Introduction}

With the rapid revolution of cloud computing, network function virtualization, and the concept of software-defined networks (SDNs) \cite{acm1} under fifth-generation (5G) umbrella, a paradigm-shift towards the sixth-generation (6G) of mobile communications is observed \cite{0}. In this era, future networks are no longer confined to conventional terrestrial cellular architectures, they are evolving towards a hybrid terrestrial-underwater-aerial-space network \cite{1,2}. Hence, the requirements of 6G are compared to those of the existing 5G networks in Table \ref{compare}. Although 6G performance indicators are not formally finalized at the time of writing, we introduce the potential target specifications from the open literature. The corresponding stringent service demands in 6G include a peak rate of 1 Tbps, system latency lower than 0.1 ms, reliability of 99.99999999$\%$, user velocity higher than 1000 km/hr, and unprecedented densities of devices per square kilometer \cite{1,sur2,sur3,4,5}. In comparison to the existing 5G system, the 6G system will require much improved energy efficiency and area spectral efficiency \cite{3,4,5}. These specifications will be able to support smooth, resilient and high-quality services in a hybrid network.  

\begin{figure*}
	\centering
	\includegraphics[width=6.3in]{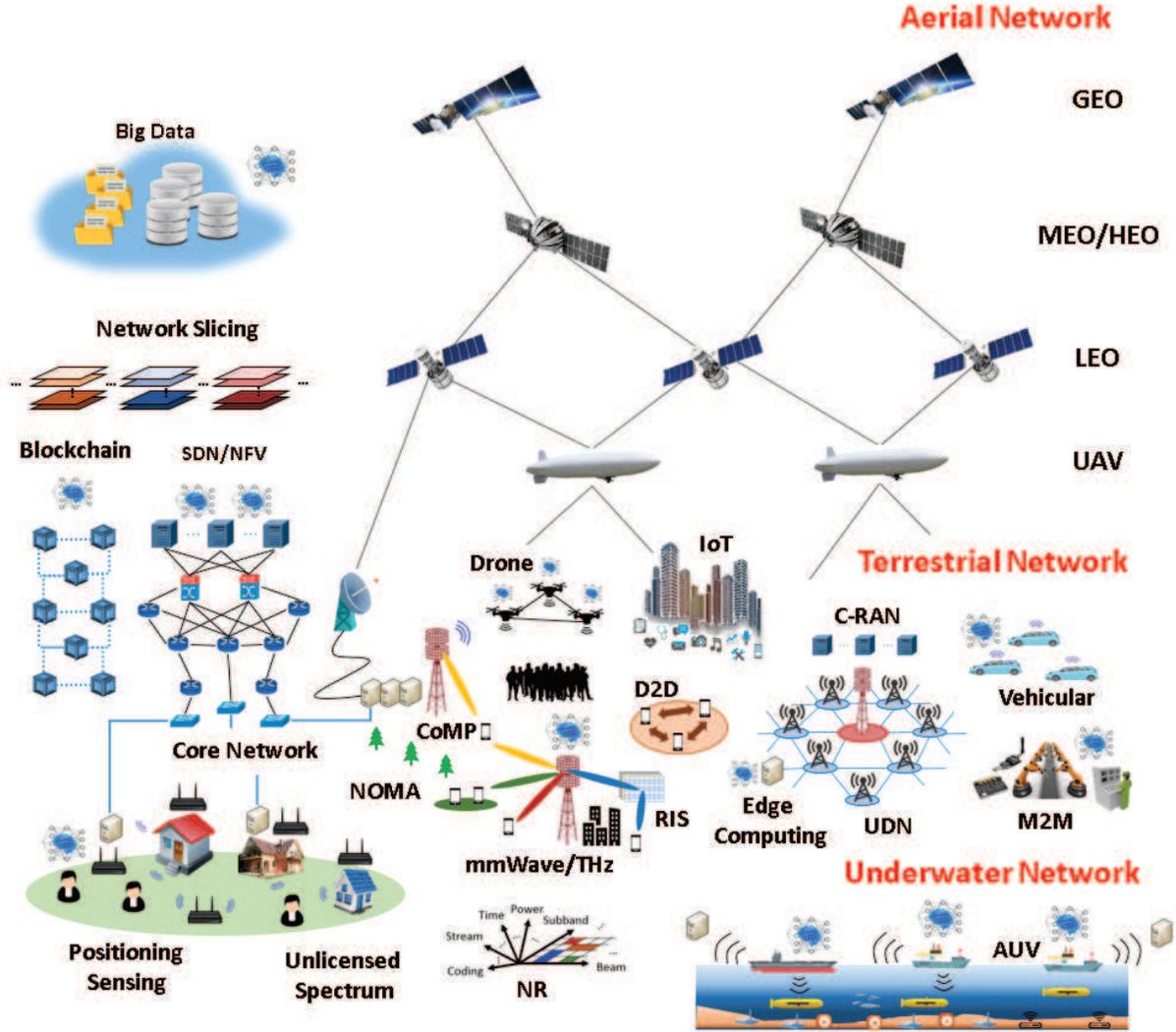}
	\caption{The architecture of AI-empowered 6G communication networking technologies includes next-generation wireless communications for aerial, terrestrial, and underwater networks. Aerial network contains GEO, MEO, LEO, UAV and drones. Terrestrial network includes V2X, C-RAN, M2M, UDN, D2D, IoT, mmWave/THz and core network, which are enabled by edge computing, RIS, NOMA, CoMP, NR, unlicensed spectrum usage, positioning and sensing, blockchain, SDN/NFV, network slicing, and big data techniques. Underwater network is formed by groups of vessels and AUVs conducting sensing and data collection missions.} 
	\label{arc}
\end{figure*}

%-------------------------------------------------
\begin{table*}
\footnotesize
\begin{center}
\caption {Key Requirement Comparison of 5G and 6G}
\renewcommand{\arraystretch}{1.1}
    \begin{tabular}{|l|l|l|}
       \hline
       & 5G & 6G \\ \hline
       Application services 
       & \tabincell{l}{\checkmark \ eMBB \\ \checkmark \ URLLC \\ \checkmark \ mMTC } 
       & \tabincell{l}{\checkmark \ eUMBB \\ \checkmark \ eURLLC \\\checkmark \ UmMTC \\\checkmark \ LDHMC \\ \checkmark \ ELPC} \\ \hline
       
       Communication network architecture  
       & \tabincell{l}{\checkmark \ 5G-NR cellular \\ \checkmark \ mmWave network} 
       & \tabincell{l}{\checkmark \ AI-empowered network \\ \checkmark \ Aerial network \\ \checkmark \ Terrestrial network \\ \checkmark \ Underwater network \\ \checkmark \ mmWave/THz network} \\ \hline
       
       Transmission spectrum usage
       &\tabincell{l}{\checkmark  \ sub-6 GHz (2.4/3.5/5 GHz) \\ \checkmark \ mmWave (28/39/60 GHz) } 
       & \tabincell{l}{\checkmark \ sub-6GHz (2.4/3.5/5 GHz) \\ \checkmark \ mmWave (28/39/60 GHz) \\ \checkmark \ THz (Above-100GHz)\\ \checkmark \ Laser\\ \checkmark \ VLC \\ \checkmark \ Non-RF} \\ \hline
       Peak data rate & 20 Gbps & 1 Tbps\\ \hline 
       
       Latency requirement & 1 ms & 0.01--0.1 ms\\ \hline
       
       Reliability demands  & 99.999 $\%$ & 99.99999999 $\%$ \\ \hline

       Connectivity density& $10^{6}$ devs/km$^{2}$ & $10^{7}$ devs/km$^{2}$  \\ \hline

       Mobility support & 500 km/hr & $\geq$ 1000 km/hr \\ \hline
       
       Area spectral efficiency compared to 5G & 1$\times$ & 10$\times$ \\ \hline
       
       Energy efficiency compared to 5G & 1$\times$ & 100$\times$ \\ \hline
    \end{tabular} \label{compare}
\end{center}
%\end{small}
\end{table*}
%------------------------------------------------
		
\begin{figure*}[!h]
	\centering
	\includegraphics[width=6.5in]{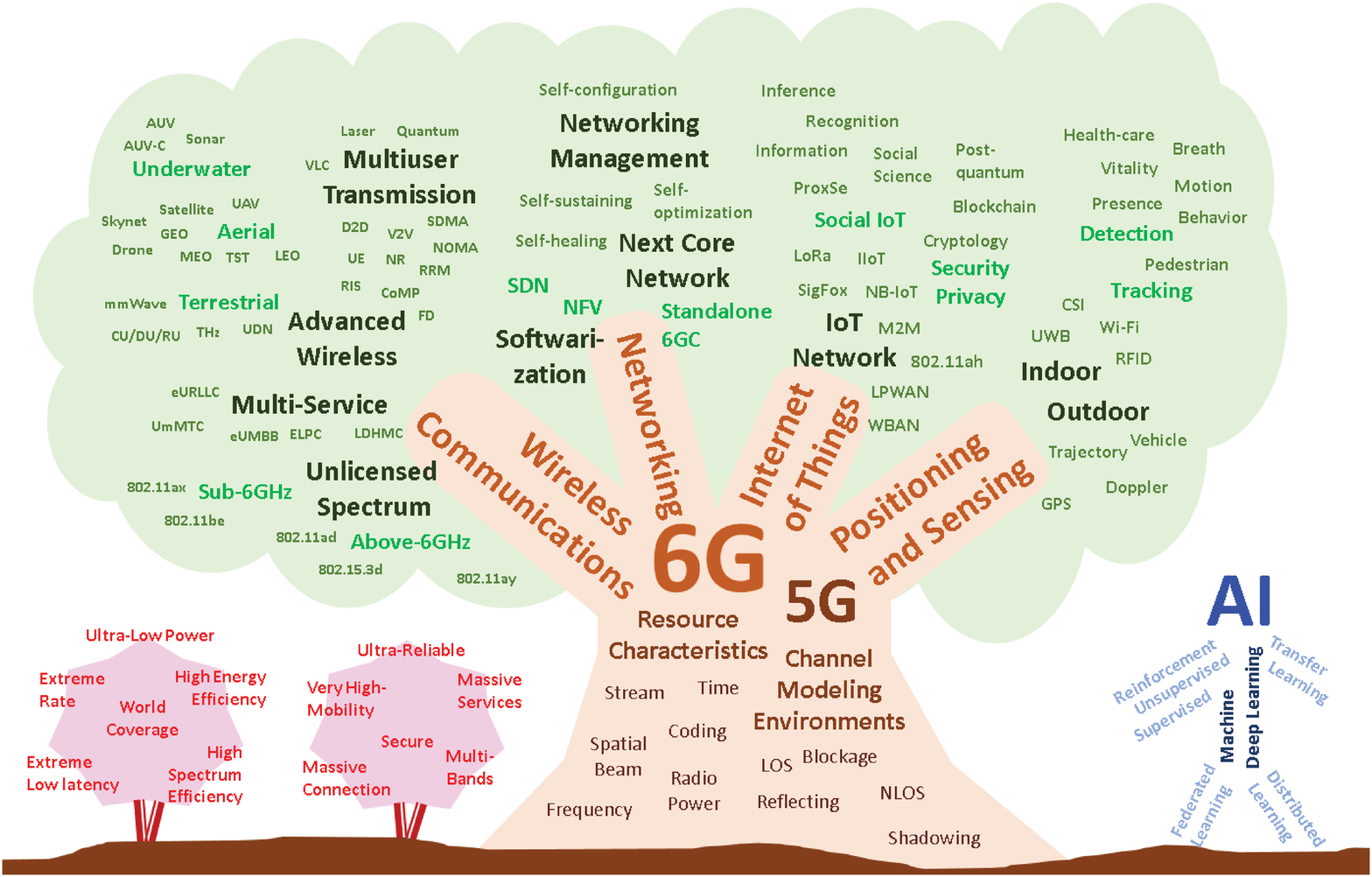}
	\caption{Nature of 6G technologies. Stemming from 5G with the root of resource characteristics and channel modeling environments, the 6G trunk includes technology branches of wireless communications, networking, Internet of Things, and positioning and sensing. The leaves and foliages of promising 6G architectures and techniques are nourished in its belonging branches from the 5G roots nourished by the human being so-called AI. Accordingly, the performance flowers are vigorously growing up shading below the lush 6G tree.} 
	\label{tree}
\end{figure*}

In this context, a whole raft of pivotal issues have to	be addressed, such as cloud storage, the underlying computing architecture, computing resource management, multimedia streaming technologies, SDN, and network function virtualization \cite{4,5}. The rapid development of artificial intelligence (AI) as a powerful optimization tool and deep learning has facilitated the solution of highly complex problems in conventional systems \cite{ai, acm4}. Advances in wireless positioning and sensing \cite{indoor} and the Internet of Things (IoT) \cite{iot, acm5} have facilitated large-scale data collection both across the industrial sectors and in the home with the prospect of supporting sophisticated new applications of 6G mobile networks. However, the massive tele-traffic forecast also leads to potential network security and privacy challenges. Accordingly, advanced information and security solutions have to be designed for supporting these novel network architectures, which are shown at a glance in Fig. \ref{arc} and will be elaborated in the following sections. Against the above backdrop, this article aims for surveying the most promising 6G research topics evolving from the 5G technologies, which are captured at a glance in Fig. \ref{tree}. The main research issues include a whole raft of \textbf{Next-Generation Architecture, Spectrum and Services}, \textbf{Next-Generation Networking}, the \textbf{Internet of Things}, \textbf{Wireless Positioning and Sensing}, as well as the \textbf{Applications of Deep Learning in 6G Networks}. The open research issues of 6G communication and networking are also summarized in Table \ref{issue}. Note that in Table \ref{issue} 6G facets 1--4 are discussed with open research issues. 
%The issues of next-generation architecture, spectrum and services include advanced wireless network, new multiuser transmission, unlicensed spectrum access, and multiple wireless services. While, the issues of next-generation networking comprise
In a nutshell, our contributions can be summarized as
follows:
\begin{itemize}
\item Relying on recent research results, we have sorted out the key performance metrics as well as five service use cases of 6G compared to 5G system. We have investigated comprehensive literature surveys for the potential promising techniques from the perspectives of architectures, networking, applications as well as scheme designs, which are extended from the current foundation of wireless and networking.

\item Furthermore, we discuss the inherent characteristics by highlighting unique and promising next-generation architecture, spectrum and services. We portray a plethora of heterogeneous architectures with integrated hybrid networks under different accessing and transmission mechanisms. The vulnerabilities of the corresponding techniques in their regions are also addressed and carefully considered for future research directions.

\item We also investigate a plentiful suite of learning-driven optimization and solutions for the above-mentioned open issues. Depending on each case and its requirement, different machine and deep learning schemes should be intelligently and favorably exploited as a remedy shown in the open literature.

\item We have demonstrated some substantial field trial performances regarding high-frequency mechanism, IoT communication and sensing as well as deep learning-driven device-free indoor positioning and sensing detection techniques.

\item Owing to future complex network scenarios with quite different requirements, we elaborate the road from single-component to multi-component Pareto-optimization principle, which differentiates conventional methods and carries out potentially encountered trade-offs among numerous factors, such as rate, bandwidth, energy, latency and complexity, etc.

\end{itemize}
Indeed, there exist other 6G research surveys and tutorials published in \cite{sur1,sur2, 4, ai, 86, sur3, sur4, sur5, sur6, sur7}. However, to the best of our knowledge, this survey has provided a more comprehensive next-generation overview of network architectures and applications as well as optimization. Additionally, our paper offers a cross-disciplinary synthesis ranging from whole network layers, which is explicitly contrasted to the existing works in Table \ref{surtable} for identifying the difference in the open literature.

\begin{table*}[t]
	\scriptsize		
	\centering
	\caption {Comparison with Available Surveys and Tutorials}
	\renewcommand{\arraystretch}{1.5}
	%\resizebox{0.7\textwidth}{!}
	\begin{tabular}{|l|c|c|c|c|c|c|c|c|c|c|c|}
		\hline
Paper
&\cite{sur1}
&\cite{sur2}
&\cite{4}
&\cite{ai}
&\cite{86}
&\cite{sur3}
&\cite{sur4}
&\cite{sur5}
&\cite{sur6}
&\cite{sur7}
& This work \\ \hline \hline
		Year 	& 2019 & 2019& 2019& 2019& 2020& 2020& 2021& 2021& 2021& 2022& 2022 \\ \hline
		Type      & Survey & Tutorial & Survey  & Tutorial & Tutorial & Survey & Survey & Survey & Survey & Survey & Survey \\ \hline 
		Netw. Architecture & & \checkmark & \checkmark  & & \checkmark & \checkmark & & \checkmark & \checkmark & \checkmark & \cmark \\ \hline
		Wireless Transmission & & \checkmark & \checkmark & \checkmark & \checkmark & \checkmark & & \checkmark & \checkmark & \checkmark & \cmark \\ \hline	
		Unlicensed Spectrum & & & & & \checkmark & \checkmark & & & & \checkmark & \cmark \\ \hline	
		Multi-Service Use Cases & & \checkmark & \checkmark & & & & & & & \checkmark & \cmark \\ \hline	
		Softwarization SDN/NFV & \checkmark & \checkmark &  & & & & & \checkmark & \checkmark & \checkmark & \cmark \\ \hline	
		IoT and Security & \checkmark & \checkmark & & \checkmark & \checkmark & \checkmark & & & \checkmark & \checkmark & \cmark \\ \hline	
		Positioning \& Sensing & \checkmark & & & & \checkmark & & \checkmark & & & & \cmark \\ \hline	
		Learning Wireless/Netw.  & \checkmark & \checkmark & \checkmark & \checkmark & \checkmark & \checkmark & \checkmark & \checkmark & \checkmark & \checkmark & \cmark \\ \hline
		Multi-Component Opt.  & & & & & \checkmark & & & & & & \cmark \\ \hline
		
	\end{tabular} \label{surtable}
\end{table*}

%-------------------------------------------------
\begin{table*}
\footnotesize
\begin{center}
\caption {Open Research Issues of 6G Communication and Networking Technologies}
\renewcommand{\arraystretch}{1.03}
    \begin{tabular}{|l|l|}
       \hline
       \specialrule{.2em}{.1em}{.1em}
       \multicolumn{2}{|c|}{\textit{\textbf{Next-Generation Architecture, Spectrum and Services}}}\\ \hline
       Advanced wireless networks
       & \tabincell{l}{\checkmark \ Interference management in hybrid 6G networks \\ \checkmark \ Dynamic spectrum management for different wireless transmissions \\\checkmark \ High-mobility handover management (e.g., vehicle, train, UAV, AUV, etc.) \\\checkmark \ Network power control and energy harvesting \\ \checkmark \ Flexible scheduling for integrated RAT} \\ \hline
       
       New multiuser transmission 
       & \tabincell{l}{\checkmark \ Enhanced RRM for diverse resources (e.g., code, space, frequency, time, etc.) \\ \checkmark \ New UM-MIMO beamforming for mmWave and THz \\ \checkmark \ Spectrum/Energy efficient techniques (e.g., NOMA, FD, CoMP, etc.) \\ \checkmark \ Advanced grant-free transmissions\\ \checkmark \ Deployment and optimization of RIS \\ \checkmark \ Non-RF techniques (e.g., laser, VLC, quantum communications, etc.)} \\ \hline
       
       Unlicensed spectrum access
	   & \tabincell{l}{\checkmark \ Interference mitigation for unlicensed spectrum accessing (e.g., LAA) \\ \checkmark \ Re-transmission mechanism at higher frequency bands \\ \checkmark \ New multiuser orthogonal contention and data transmission schemes\\ \checkmark \ Advanced unlicensed 60 GHz beamforming design for 802.11ad/ay} \\ \hline
       
       Multiple wireless services
       & \tabincell{l}{\checkmark \ High performance hybrid services (e.g., URLLC-eMBB) \\
\checkmark  \ Front-end resource allocation and hybrid numerology optimization \\ \checkmark \ Management of new 6G services of LDHMC and ELPC} \\ 
       
       \specialrule{.2em}{.1em}{.1em}
       \multicolumn{2}{|c|}{\textit{\textbf{Next-Generation Networking}}}\\ \hline
       
       Network softwarization 
       & \tabincell{l}{\checkmark \ Advanced automatic network traffic optimization and service management \\
\checkmark  \ Flexible and cost-effective network function deployment \\ \checkmark \ New NFV management and orchestration} \\ \hline
       
       Next-generation core
       & \tabincell{l}{\checkmark \ Next-generation core virtualization \\
\checkmark  \ QoS-guaranteed virtual networks} \\ \hline

       Mobile network management
       & \tabincell{l}{\checkmark \ Resource management for mobile cloud and edge computing \\
\checkmark  \ Advanced SDN/NFV-enabled SON} \\ 

\specialrule{.2em}{.1em}{.1em}
\multicolumn{2}{|c|}{\textit{\textbf{Internet of Things}}}\\ \hline
      
       IoT networks
       & \tabincell{l}{\checkmark \ Efficient mechanisms for sensing and data collection and upload \\ \checkmark  \ Simultaneous operation among different IoT protocols \\ \checkmark \ Advanced power preservation, interference management and synchronization \\ \checkmark \ Cloud resource management and big data storage and processing \\ \checkmark \ Advanced social IoT network structure for information dissemination and recognition} \\ \hline

		Vehicular networks
       & \tabincell{l}{\checkmark \ Wireless channel characteristic and resource management of V2X\\ \checkmark  \ Heterogeneity management over different V2X protocols and interfaces \\ \checkmark \ Advanced optimization of joint sensing, control and communications \\ \checkmark \ Efficient and effective vehicular routing and trajectory design \\ \checkmark \ Congestion control and secured and reliable IoT-V2X} \\ \hline

       Security and Privacy
       & \tabincell{l}{\checkmark  \ Next-generation quantum and post-quantum cryptology \\
\checkmark  \ Implementations/Applications of multifunctional security and privacy techniques \\ \checkmark \ Blockchain on advanced data security and system operation 
\\ \checkmark \ Enhancement in physical layer security
\\ \checkmark \ Privacy-aware strategies in smart services and cloud/edge networks} \\

\specialrule{.2em}{.1em}{.1em}
\multicolumn{2}{|c|}{\textit{\textbf{Wireless Positioning and Sensing}}}\\ \hline
       
       Wireless positioning and sensing
       & \tabincell{l}{\checkmark \ Flexible, robust and high-precision trajectory tracking \\ \checkmark \ Doppler shift of high-speed and long-distance outdoor positioning \\
\checkmark  \ Device-free CSI-based positioning, sensing, and detection \\ \checkmark \ Fine-grained positioning and detection in extreme environments \\ \checkmark \ Fine-grained positioning and detection in extreme environments \\ \checkmark \ Multi-scale human behavior and vitality detection} \\ 

\specialrule{.2em}{.1em}{.1em} \hline

    \end{tabular} \label{issue}
\end{center}
%\end{small}
\end{table*}
%------------------------------------------------
	
\section{Facet 1: Next-Generation Architecture, Spectrum and Services}

\subsection{Advanced Wireless Network Architecture and Technology}
	The forthcoming 6G wireless network is expected to evolve beyond the conventional terrestrial cellular network by additionally including underwater, aerial and satellite communication networks, forming a vertical 3D network (so-called 3DNet or SkyNet) \cite{2, 3d} as seen in the stylized illustration of Fig. \ref{arc}. Accordingly, the management of these emerging heterogeneous vertical/horizontal massive ultra-dense networks (UDN) becomes one of the key research topics. The conventional cloud radio access network (C-RAN) \cite{HanzoCRAN} relies on fully centralized network functions, computations, decisions and operations in the central cloud, which is insufficiently flexible for future networks. Accordingly, based on the associated network functions, 6G \textit{terrestrial communication networks} partition the traditional base stations (BSs) of the C-RAN into the central unit (CU), distributed units (DUs), and radio units (RUs). This partitioning requires flexible radio access technology (RAT) \cite{6}. The powerful CU in the cloud has substantial computation and data storage capabilities managed by the higher network layers, whilst the network functions of the lower layers are deployed within multiple DUs at the edge. The RUs are responsible for signal transmission and reception, while the networking policy is formulated at the DUs and CU. Furthermore, instead of using traditional costly wired links, the fronthaul and backhaul of flexible CU/DU/RU architectures may rely on high-speed millimeter wave (mmWave) and terahertz (THz) techniques \cite{7}. Similarly, both the user equipment (UE) and sensors are supported by high-speed mmWave THz radio links.

\begin{itemize}
	\item \textbf{mmWave Transmission}: As a benefit of its ample bandwidth, a mmWave system typically operating at 28/39/60 GHz is capable of supporting Gbps-level transmission \cite{mm1,mm2,mm100}. However, this is achieved at a high pathloss and sensitivity to blockages. As a potential remedy, beamforming relying on high-gain beams may be harnessed for mitigating the pathloss \cite{mm1}.
	
	\item \textbf{THz Transmission}: The THz band represents the carrier frequencies spanning from 0.1 to 1 THz, which has even wider bandwidth resources than the mmWave band. Hence, it is potentially capable of supporting Tbps transmission speeds \cite{1,7}. However, it suffers from significantly higher pathloss than mmWave carriers due to severe molecular absorption. Therefore, it requires massive antennas to support so-called THz-oriented pencil-beams \cite{t1,t2}. In this context, THz beam-alignment is an excessively challenging task, especially in the face of mobility in a short-distance THz communications. Moreover, directional THz-enabled cognitive radio (CR) aided transceivers are capable of dynamically exploiting the slivers of unoccupied spectrum for improving the area spectral efficiency.
\end{itemize}

	The 6G architecture of \textit{underwater communication networks} has the task of supporting autonomous underwater vehicles (AUVs) and AUV controllers (AUVCs) \cite{9}, which can be harnessed in diverse scenarios, including underwater air crash investigations, military applications, and deep sea exploration. The AUV-based network supports multiple AUVs and multiple AUVCs equipped with sensors and sonar/camera systems for collaborative navigation, localization, and object tracking. However, underwater communications rely on ultra-low frequencies, which are affected by water flow, the Doppler effect of ships, environmental noise, and vortex-induced water vibration. The mitigation of these phenomena requires substantial further research in 6G underwater networks.
	
	\textit{Aerial communications} in 6G rely on drones or unmanned aerial vehicles (UAVs) and low/high altitude platforms (LAPs/HAPs). Furthermore, the 6G satellite network contains several layers in Fig. \ref{arc}, including the low earth orbit (LEO) satellites below 2000 kilometers, medium earth orbit (MEO), high earth orbit (HEO), and geostationary earth orbit (GEO) satellites at 36,000 kilometers \cite{2,10}. Due to the long transmission distance from the satellite to ground, high transmit power is required for mitigating the pathloss and specialized terrestrial-space terminals (TSTs) have to be used. For efficiently collecting information, multiple UAVs and multiple satellites may cooperatively transfer their data forming a heterogeneous integrated ground-air-space (IGAS) network \cite{u0}. However, numerous mobility-related factors should be taken into account, such as the roll/pitch/yaw movement of UAVs and the high-velocity orbiting of LEO and MEO satellites are challenging issues to be tackled \cite{u1,u2}. But again, the UAVs and satellites are capable of substantially improving the coverage quality \cite{u1,u2}.

	As an evolution from conventional BS-centric networking, user-centric cell-free networks \cite{11} have become popular, which judiciously allocate the network's resources according to the specific quality of service (QoS) requirements of the UE. As a benefit of this user-centric philosophy, amorphous coverage areas are created by assigning the access points inhomogeneously by matching their density to the non-uniform user-density. Hence, they exhibit excellent load-balancing capability. The network determines its resource-allocation strictly based on the QoS requirements \cite{HanzoQoS}. However, achieving this ambitious design objective, while handling diverse cell-sizes, ranging from small cells (SCs) to femtocells, picocells and the emerging nanocells \cite{11-2}, requires substantial further research in the context of 6G networks. To elaborate a little further, a host of sophisticated interference management, dynamic channel allocation, high-mobility handover, packet admission control, power control, and scheduling have to be investigated. Furthermore, the nodes operating in remote areas, where no electricity is available have to rely on advanced energy harvesting and wireless power transfer in next-generation wireless communications \cite{12, 12-1, swipt-hanzo}.

\subsection{New Multiuser Transmission Schemes}	
	Conventional transmission techniques tend to focus on improving their resource efficiency in time, frequency, and spatial domains. The 6G new radio (NR) will further extend these techniques to the mmWave and THz bands \cite{7}, whilst relying on ultra-massive multiple input multiple output (UM-MIMO) beamforming techniques \cite{13, HanzoMIMOOFDM}. By relying on these techniques, multiple beams can be generated to serve numerous users in different directions at diverse QoS requirements. However, sophisticated power control, interference management and radio resource management (RRM) are required. Moreover, in order to improve the spectral- versus energy-efficiency trade-off, we can superpose all signals in the time, frequency and spatial resource slots. The corresponding techniques include non-orthogonal multiple access (NOMA) \cite{14}, 3D beamforming, full-duplex (FD) \cite{15}, coordinated multi-point (CoMP) transmission/reception \cite{16}, and rate splitting multiple access (RSMA) \cite{rsma1}.

\begin{figure*}[!t]
\centering
\subfigure[]{\includegraphics[width=1.5in]{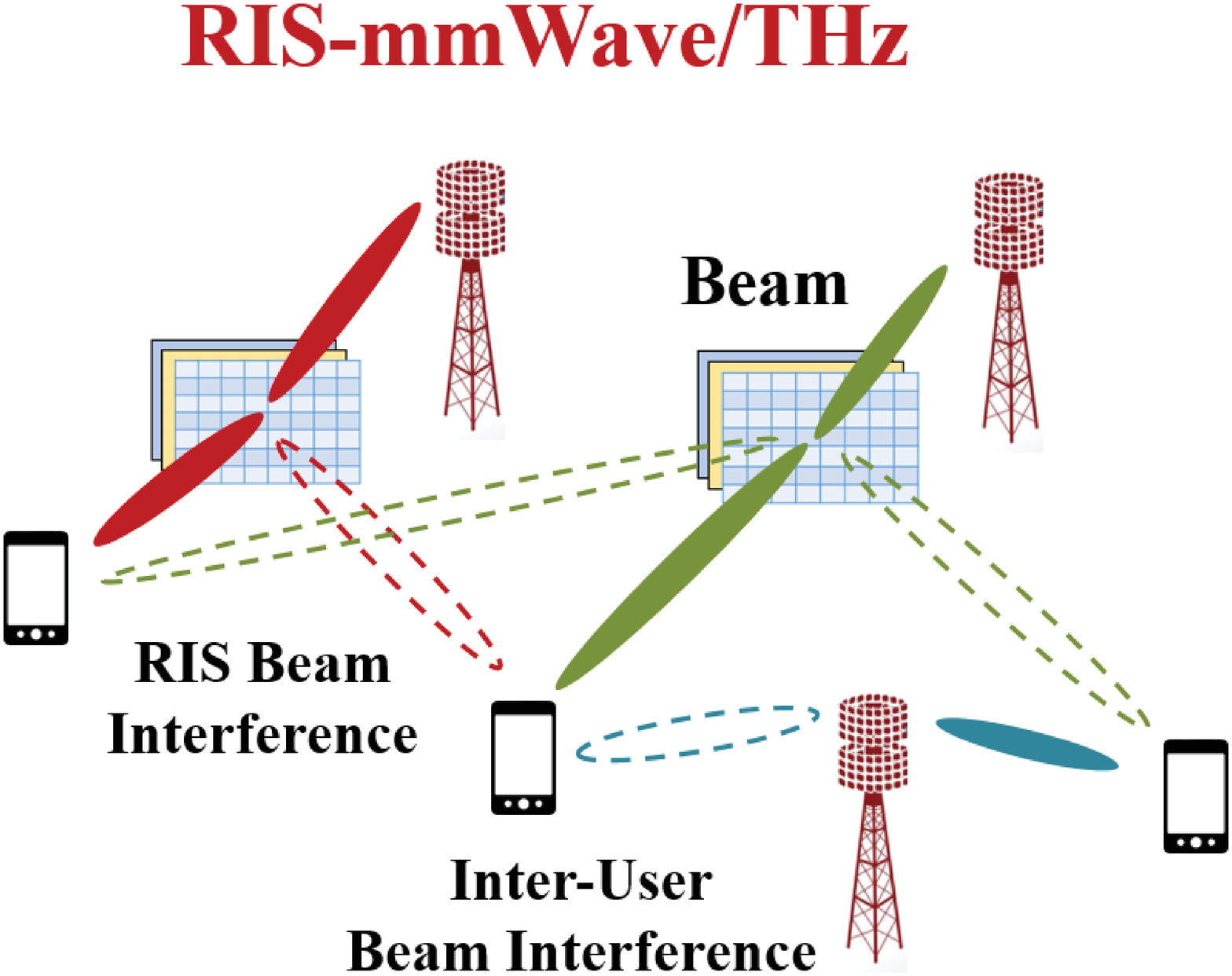} \label{RIS1}}
\subfigure[]{\includegraphics[width=1.5in]{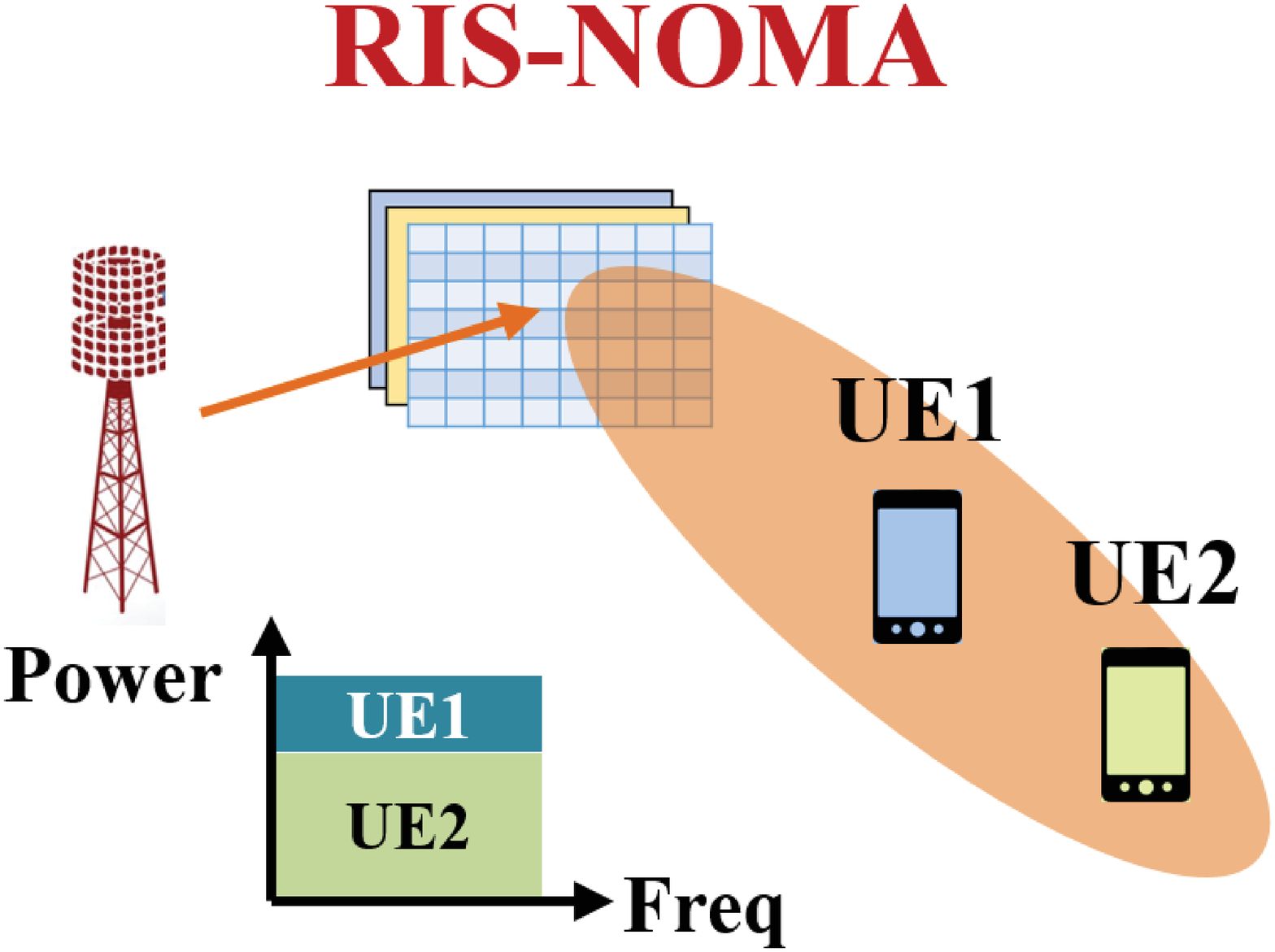} \label{RIS2}}
\subfigure[]{\includegraphics[width=1.5in]{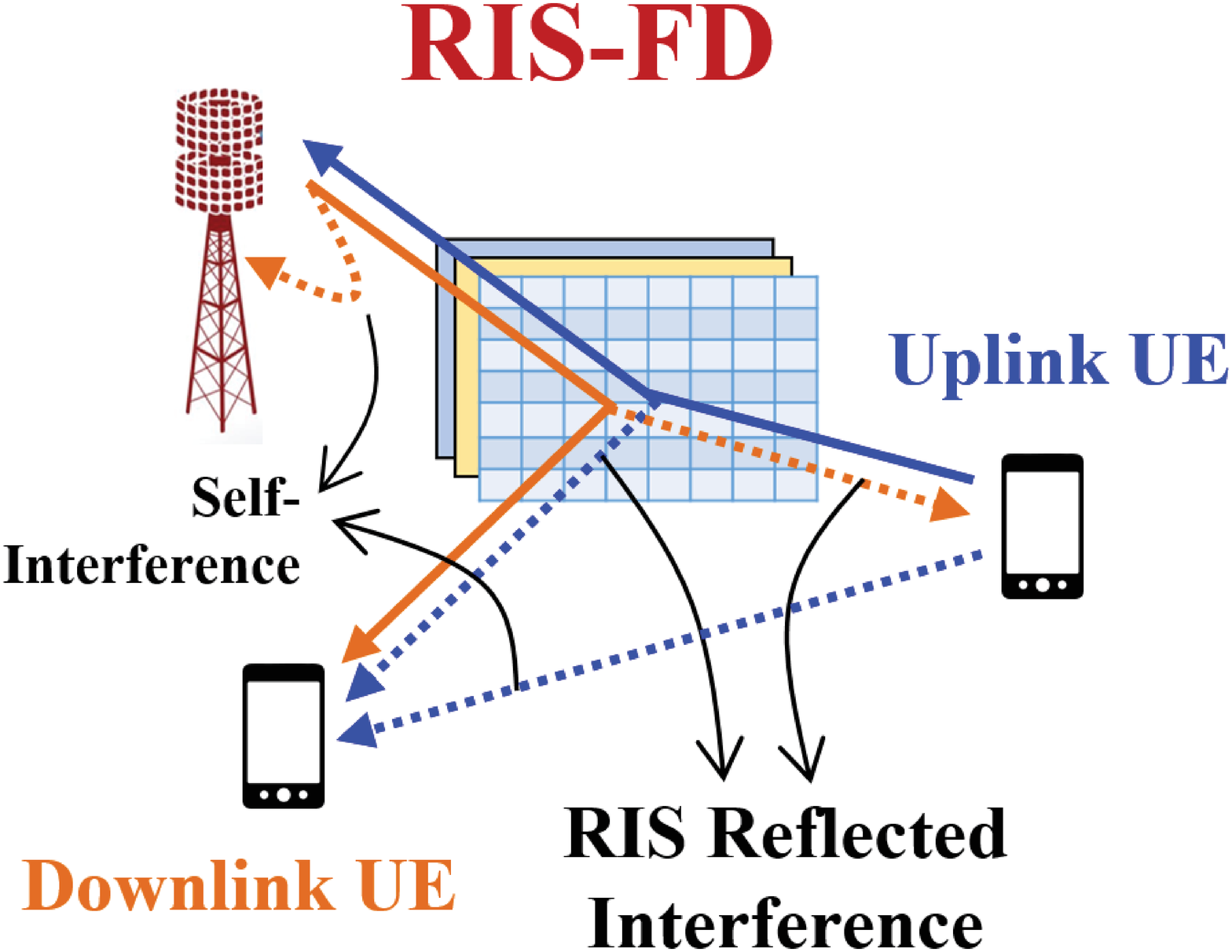} \label{RIS3}}
\subfigure[]{\includegraphics[width=1.5in]{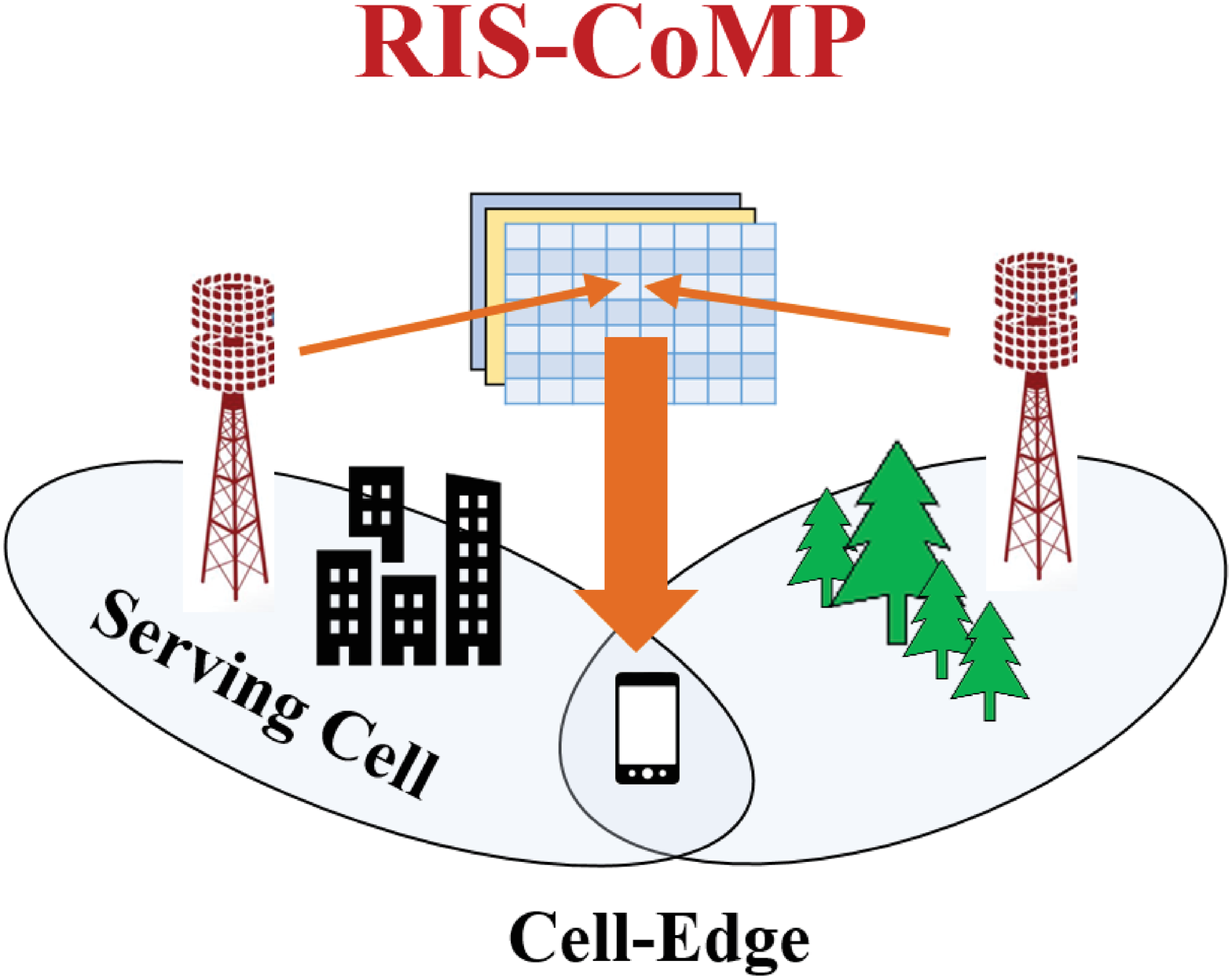} \label{RIS4}}\\
\subfigure[]{\includegraphics[width=3.9in]{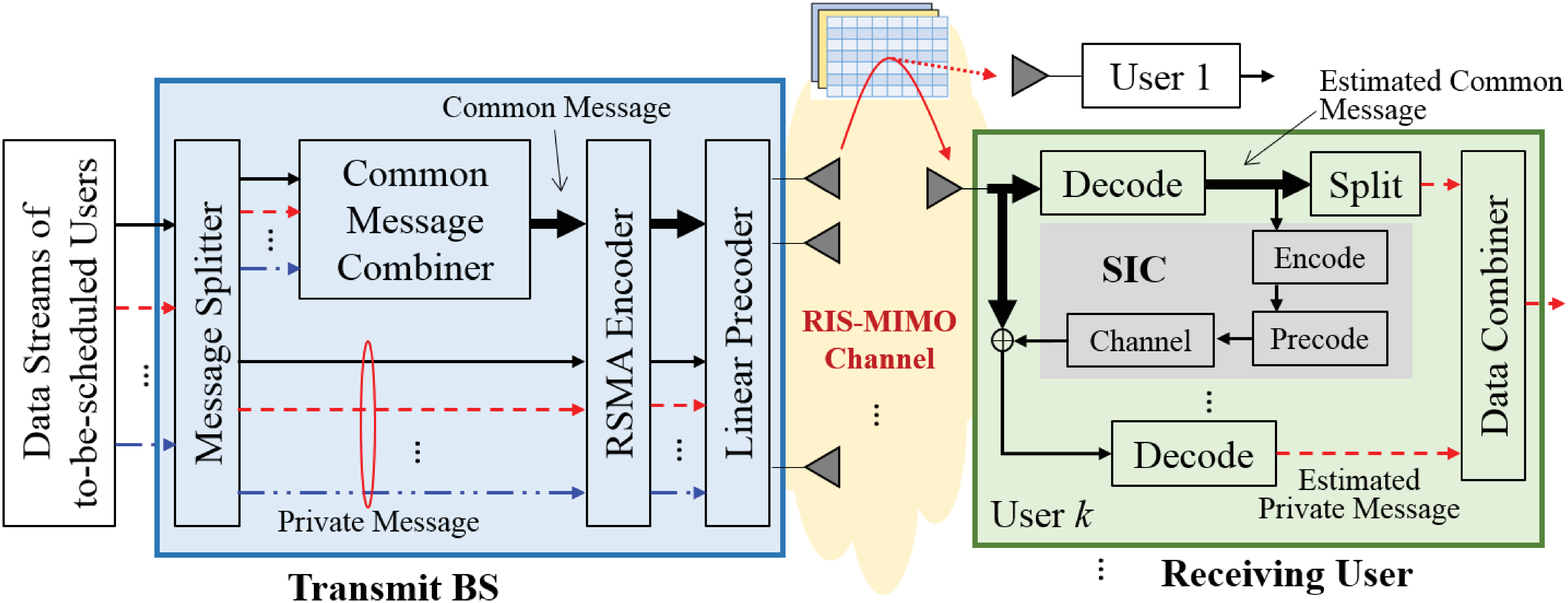} \label{risrsma}}
\subfigure[]{\includegraphics[width=2.1in]{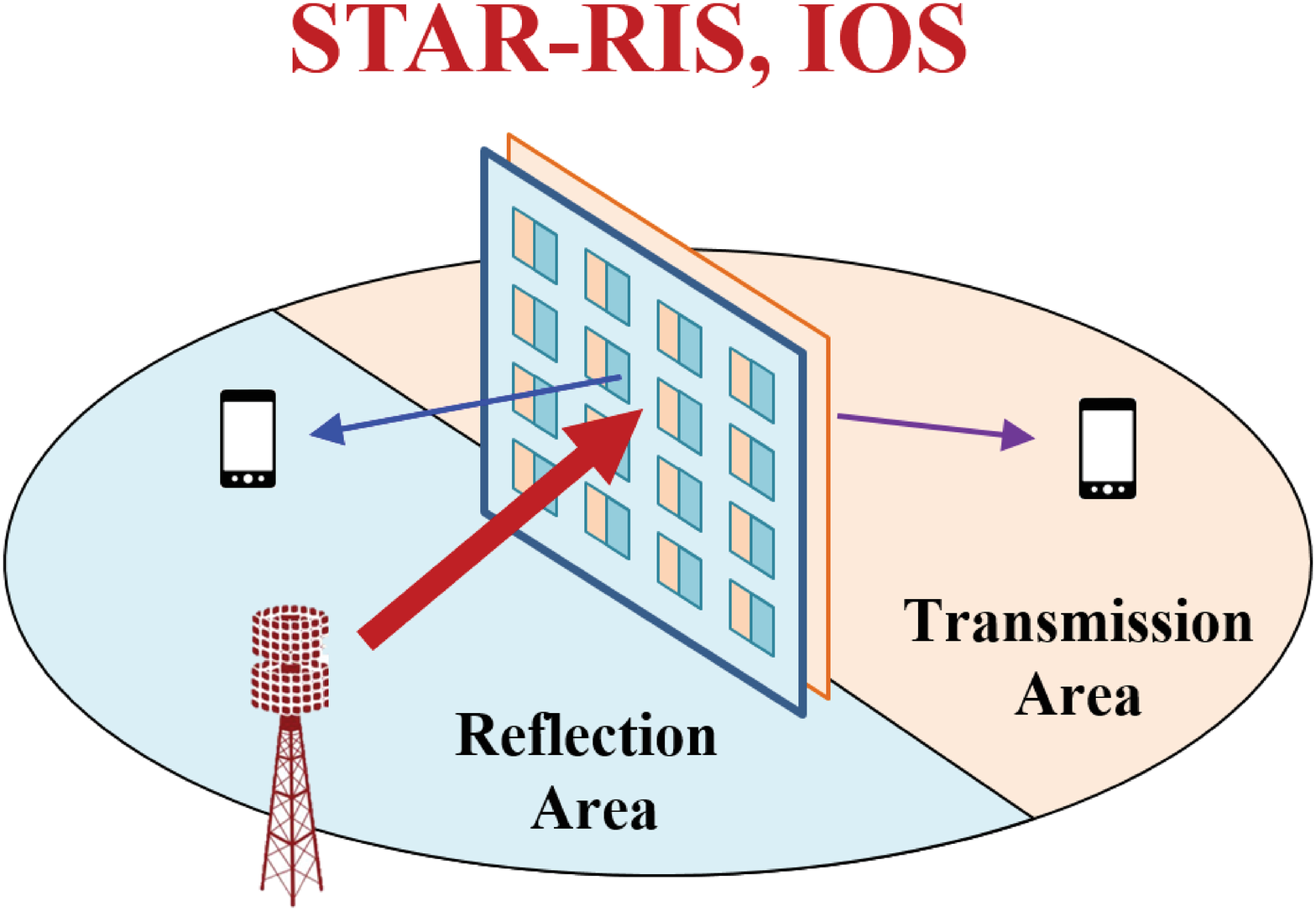} \label{risstar}}
\caption{The 6G RIS-empowered wireless network of (a) RIS-empowerd mmWave/THz (b) RIS-NOMA (c) RIS-FD and (d) RIS-CoMP transmissions as well as emerging architecture of (e) RSMA and (f) STAR-RIS/IOS.} \label{RISplot}
\end{figure*}

\begin{itemize}
\item \textbf{NOMA}: The transmitter will multiplex several desired signals to a single resource slot of certain resource domains, whilst the receiver carries out successive interference cancellation (SIC) for mitigating the interference imposed by the signals belonging to other users. In the most of popular power-domain NOMA two or more users' signals are superimposed at a given power and SIC is used to detect the strongest signal, while considering the weaker signals as interference. The remodulated signal is then deducted from the aggregate signal, leaving the clean/weaker signal behind \cite{n1,n2,c4}. Similar SIC-aided procedures are used also for code-domain NOMA. A whole plethora of other NOMA solutions can be found in \cite{HanzoNOMA1,HanzoNOMA2}. However, it becomes a potential challenge to design advanced NOMA techniques in terms of integrating the diverse resource domains of mmWave/THz systems \cite{n3,n4}, cancelling the interference in UDNs, or enhancing the coverage area of UAV and satellite networks \cite{n5,n6}.

\item \textbf{3D Beamforming}: This concept emerges from that of sectorized antennas aiming for serving users roaming at different angles with the aid of sophisticated beamforming techniques and antenna configurations. Since mmWave carriers suffer from excessive pathloss, the employment of high-gain beamforming is critical. High-gain beamforming is even more crucial for THz carriers relying on pencil beams. Furthermore, it is of salient importance to frugally manage the limited 3D resources for supporting diverse QoS requirements under IGAS networks.

\item \textbf{FD}: This technique allows simultaneous uplink and downlink transmission within a single timeslot at the same frequency \cite{15, c5}. However, the interference imposed by the high transmit power on the low received power is a critical issue, which requires advanced self-interference mitigation techniques. With the emergence of powerful new 6G architectures and technologies, there is an opportunity for FD solutions to increase the spectral efficiency by employing mmWave/THz UM-MIMO beamforming in both terrestrial and aerial networks \cite{f1,f2,f3}.

\item \textbf{CoMP}: This technique supports simultaneous transmissions from multiple BSs to a single receiver \cite{c1}. Furthermore, as a benefit of mmWave/THz beamforming, UM-MIMO CoMP is capable of increasing the network's throughput \cite{c2,c3}. Nonetheless, there are substantial challenges in the way of large-scale CoMP roll-out, such as the related synchronization issues, because the receiver can only be perfectly synchronized with a single BS.

\item \textbf{RSMA}: As depicted in Fig. \ref{risrsma}, considering the base stations having for example $M$ transmit antennas and $K$ users -- each relying on a single receiver antenna -- communicate under the assumption that in addition to the private and confidential messages destined for the individual users, there are also common messages to be received by all of the users. The terminology of rate-splitting of messages implies that the downlink stream is partitioned into $K$ segments for the $K$ users, where each user's message contains both a common and a private message segments. The $K$ private messages of the individual users are then jointly transmitted with the common message of all users. Both the common and private messages are transmitted in the downlink by the BS having $M$ downlink transmit antennas, but naturally, the antenna-array weights used for the transmit precoding (TPC) private messages depend on the individual user positions, while those used for the common part are trained for reaching all users \cite{rsma1, rsma2, rsma3, rsma4}.

At the users, the common message is detected first by assuming that the private messages are unknown and hence they can only by treated as additional noise. In the next processing step, we aim for cancelling the interference imposed by the common message on the composite received signal using SIC. This is achieved by first remodulating the common message detected as well as applying the TPC to its modulated version and then subtracting the result from the composite received signal. This leaves the superimposed private messages behind. Now each user has to extract his/her own private message following a similar SIC process as outlined above. Explicitly, each user detects his/her own private signal by treating all the other users' signals as noise. This is because the TPC weights of the other users are unknown and hence their interference cannot be cancelled. Suffice to say that the above rudimentary portrayal of the RSMA philosophy relies on a number of simplifying assumptions, which are eliminated in the detailed treatises of \cite{rsma-hanzo1, rsma-hanzo2}.

%By relying the same technique of SIC in NOMA, RSMA \cite{rsma1, rsma2, rsma3, rsma4} as a promising PHY transmission technique treats residual interference as noise while tackling the others through SIC for UM-MIMO systems. The significant feature of RSMA is that the BS transfers part of the information in a commonly encoded shared signal stream and decodes partial inter-user interference under interfering channels. RSMA has addressed the drawbacks of orthogonal transmissions and NOMA especially under imperfect CSI measurement. Benefited by SIC, RSMA fully exploits the multiplexing gain in order to enhance the deployment and configuration robustness, higher rate, escalating serving users as well as a larger scale of multi-antenna arrays. However, the open challenges of RSMA consist of receiver SIC implementation, multiuser pairing, joint user and message scheduling, as well as multi-carrier and novel waveform designs.

\end{itemize}

Reconfigurable intelligent surfaces (RISs) also constitute promising techniques for extending the coverage area, reducing the power consumption, and enhancing the data rates \cite{17,r1,r2,r3,r4}. The RIS is composed of numerous metamaterial elements, which can reflect the received waves, while adjusting their phases without complex signal processing \cite{r4} as detailed below.

\begin{itemize}
\item \textbf{RIS-Empowerd mmWave/THz Transmissions}: The fixed BS infrastructure can beamform its mmWave/THz signals to the RIS, which may reflect them to arbitrary transmit directions \cite{r5,r6,r7}, as demonstrated in Fig. \ref{RIS1}. Blocking the line-of-sight (LOS) paths of mmWave/THz carriers may be circumvented with the aid of RISs. However, they create extra interference, which has to be carefully managed. In this context, it is imperative to jointly design the active beamforming at the BS and the passive phase shift based beamforming at the RIS in order to meet different requirements.

\item \textbf{RIS-Empowerd Multiuser Transmissions}:

(1) \textit{\bf RIS-NOMA}: Again, multiuser NOMA schemes impose extra interference due to the superposed signals of the 3D resource domains. The RIS deployment shown in Fig. \ref{RIS2} has the potential of generating specific channel features for readily distinguishing the overlapped NOMA signals. Moreover, RIS-NOMA \cite{r1,r8,r9,r10} is capable of extending the coverage area to provide services for distant cell-edge users. However, the joint design of the different resource domains of NOMA and RIS constitutes a wide open research issue.

(2) \textit{\bf RIS-FD}: As for FD transmission, RISs are capable of adjusting their phase shifts to cancel or alleviate the self-interference of FD \cite{r11}, where the uplink and downlink signals may become orthogonal in terms of their directions as demonstrated in Fig. \ref{RIS3}. The RISs are intrinsically operated at FD which directly reflect arbitrary incident signals. We note that the FD transmission here indicates the co-existence of two transmission directions in the wireless network, including downlink signals from BS to UE and simultaneous uplink access in a reverse direction. However, additional interference emerges from the RISs, which should be jointly considered in RIS-FD design \cite{r12,r13}. The joint design of FD and RIS configurations may achieve potentially higher area spectral efficiency than conventional FD operating without RIS assistance.

(3) \textit{\bf RIS-CoMP}: Conventional CoMP aims for improving the low signal quality of cell-edge users. This is achieved by turning the harmful interference into useful source of desired signal energy with aid of RIS-CoMP \cite{r14}, as seen in Fig. \ref{RIS4}. As a benefit, the transmitter is capable of dissipating less power than conventional CoMP while still meeting the tele-traffic demands. Accordingly, CoMP-RIS should be jointly designed for improving the area spectrum/energy efficiency. However, there are numerous open problems, including their BS backbone bottlenecks, channel estimation and synchronization, just to mention a few. Again, it is not possible to perfectly synchronize a UE with more than one BS.

\item \textbf{Simultaneous Transmitting and Reflecting RISs (STAR-RIS)}: An impediment of the conventional RIS solutions is that the transmitter and the user have to be within the same 180-degree half-plane, rather than roaming across the entire 360-degree full plane. By contrast, the STAR-RIS architecture as illustrated in Fig. \ref{risstar}, or termed as intelligent omni-surface (IOS), allows full-plane coverage by potentially harnessing full reflection, full transmission, as well as simultaneous transmission and reflection \cite{star-ris1, star-ris2}. These modes were discussed in \cite{star-ris1} with special emphasis on NTT DOCOMO's prototype. There are three different principles governing their operations, namely the so-called energy-splitting, partitioning and time-switching types, which have their different pros and cons \cite{star-ris1}. In the energy-splitting mode the signal impinging upon an element is partially reflected and transmitted. By contrast, the partitioning type may be viewed as having a reflection-only and transmission-only segment of reduced sizes. Finally, the time-switching type is capable of switching the reflective elements between the transmit and reflect modes. There is a huge variety of compelling applications scenarios, such as STAR-RIS-NOMA \cite{star-ris3}, STAR-RIS-CoMP \cite{star-ris4} and multi-STAR-RIS deployment as well as AI-assisted STAR-RIS \cite{star-ris5}, which require further exploration by the research community.

%Different from conventional RISs, a brand-new STAR-RIS, or so-called intelligent omni-surface (IOS) is capable of simultaneously transmitting and reflecting the desired signals in full 360-degree coverage \cite{star-ris1}. Note that the transmission function is activated by penetration property of surface but with the beamformed signal power shared with reflection elements of RISs \cite{star-ris2}. The practical protocols include energy splitting, mode switching and time switching based STAR-RIS \cite{star-ris1}. Therefore, STAR-RIS becomes more flexible to be deployed anytime and anywhere, performing through-the-wall service inside and outside the building. Emerging novel architectures leverage existing physical layer techniques, such as STAR-RIS-NOMA \cite{star-ris3} and STAR-RIS-CoMP \cite{star-ris4}, which provide not only coverage but comparably stronger signal strength. However, there induce open practical challenges, including joint STAR beamformer optimization, full-coverage physical layer security, IOS circuiting and manufacturing, as well as multi-STAR-RIS deployment and control, which should be addressed in future works.

\end{itemize}

Furthermore, from an air-interface and transmission framing perspective, both multi-numerology, as well as mini-slot based and grant-free transmissions potentially make the systems more flexible in terms of reusing the time/frequency and spatial domain radio resources. However, several possible issues arise in the emerging 6G communication systems, which are elaborated as follows.

\begin{itemize}
\item \textbf{Multi-domain numerology}: Given the wide range of diverse applications and services, the multi-domain numerology defined in 5G new radio (5G-NR) enables flexible configuration of the time and frequency slots, where several resource elements can be specifically configured for meeting the QoS requirements encountered \cite{nr0,nr1}. However, this flexibility is attained at a potentially severe inter-domain interference \cite{nr2,nr3,nr4}. With the emergence of 6G networks, advanced multi-domain numerology is required for defining a common air interface for supporting hybrid mmWave/THz and multiuser transmission schemes under the integrated 3D UDN philosophy.

\item \textbf{Mini-slots}: As a similar concept to that of multi-domain numerology, the advanced philosophy of reserving mini-slots within a timeslot for prioritized latency-aware or reliability-aware services has emerged \cite{ms1}. However, the provision of mini-slots is challenging due to the associated dynamic configuration required by the multi-domain numerology aided multiuser transmission schemes. Furthermore, how to strike a beneficial compromise between the existing services and the emerging 6G applications and scenarios is a substantial open challenge.

\item \textbf{Grant-free transmission}: Conventional grant-based uplink transmission imposes high latency owing to performing four-phase handshakes relying on access request, access grant, transmission, and acknowledgement \cite{g1}. By contrast, grant-free transmission facilitates direct uplink transmission under a simplified two-phase procedure of transmission and acknowledgement \cite{g2,g3,g4}. Despite the gradual maturing of this subject area, the conception of low-overhead grant-free transmission for hybrid multi-domain numerology and mini-slots requires dedicated community-effort. This is particularly urgent in the area of joint user-activity and channel estimation, as well as iterative synchronization and data detection \cite{HanzoGrant1, HanzoGrant2} for multiuser network architectures.

%\item \textbf{Orthogonal Time-Frequency Space (OTFS) Modulation}: As an emerging novel modulation technique, OTFS unleashes the benefits and its potential in high mobility networks in vehicular, non-terrestrial and underwater acoustic networks with higher frequencies of mmWave/THz bands \cite{otfs1,otfs2}. In convention, OFDM systems perform time-division and frequency-Division multiplexing, which require additional compensation or filtering owing to distortion from mobility. By contrast, OTFS is capable of providing Doppler resilience in time-variant channel models through modulating signals in delay-Doppler domain \cite{otfs3,otfs4,otfs5}. Compared to OFDM and the existing modulations, OTFS achieves comparably lower peak-to-average power ratio, less idle time of guard-intervals, and more robust against the carrier frequency offsets. However, OTFS is still in infancy with diverse critical challenges to be addressed, such as channel estimation, detection, coding, scalable multiple access, as well as integration of sensing and communications.

\item \textbf{From OFDM to OTFS}: As part of the evolution of wireless communication through five generations the system capabilities have improved by orders of magnitude, in particular the achievable bit rate. The corresponding symbol durations have been reduced commensurately, which results in ever more dispersive channels requiring more powerful higher-order channel equalizers. As part of this trend, it became clear that using single-tap frequency-domain equalization as in OFDM is a more attractive solution for high-rate systems operating in dispersive channels than using excessive-order time-domain equalizers. As another dominant trend of the same era, the vehicular velocity has also been escalating and so did the carrier frequency, since high-rate high-bandwidth systems can only be accommodated at high carrier frequencies, where unused bandwidth is still available \cite{otfs1,otfs2}. This trend heralded the era of high-Doppler systems. Against this backdrop it is clear that a fundamental understanding of wireless propagation relying on the family of Bello-functions \cite{otfsbook} is of pivotal significance. This is particularly in the context of high-velocity UAV, aeroplane and satellite communications, which is likely to become part of the space-air-ground integrated network (SAGIN) networking concept \cite{sagin}, or referred to as IGAS of the emerging 6G systems.

In the associated high-mobility and high-Doppler contexts it becomes attractive to carry out the associated signal processing in the so-called delay-Doppler domain \cite{otfsbook}, rather than relying on the classic time-frequency-domain OFDM principles. This avenue of thought leads to the concept of orthogonal time-frequency space (OTFS) modulation transceivers \cite{otfs3,otfs4,otfs5}, as detailed below. In simple tangible terms a linear time-invariant (LTI) system having a time-invariant channel impulse response (CIR) has an infinite coherence time, where each CIR tap remains constant versus time. Clearly, this is a dispersive CIR and its Fourier transform gives the frequency-domain channel transfer function (FDCHTF). But again, in practice CIR taps tend to fluctuate, even the receiver is stationary owing to the movements of people and objects, hence resulting in linear time-variant (LTV) channels which impose frequency shifts due to the Doppler effect, yielding frequency-domain dispersion. Recall that based on the Fourier transform, a non-dispersive Dirac-delta CIR results in a flat FDCHTF, which time-dispersive CIRs result is frequency-selective FDCHTF. By the same token, high-Doppler frequency-dispersive channels are time-selective and in reality the LTV channels of high-mobility scenarios are typically both time- and frequency-dispersive upon encountering long-delay CIRs and high-velocity, high-Doppler propagation scenarios. The classic OFDM systems tend to use adaptive bit-loading of the subcarriers of the 2D time-frequency plane, which have found their way into numerous systems, including the 4G and 5G systems. Instead, the more recent OTFS technique relies on the above-mentioned delay-Doppler (DD) domain and it is shown to be capable of outperforming OFDM, especially in high-Doppler SAGIN applications \cite{hanzo-otfs1}. As a benefit of the sparse and quasi-stationary nature of the DD-domain, convenient low-overhead DD-domain channel estimation becomes possible \cite{hanzo-otfs2}, but there are numerous open problems to be addressed by future research in \cite{hanzo-otfs1, hanzo-otfs3, hanzo-otfs4}, such as the choice of the most appropriate near-capacity channel codes and multiple access techniques, just to name a few.

\end{itemize}

In contrast to traditional radio transmission, advanced next-generation technologies will include non-radio frequency (Non-RF) solutions relying on laser based optical communications, visible light communications (VLC) and quantum communications \cite{18,vlc,q1,q2}. The pertinent research issues consist of power control and modulation design for laser based optical wireless communications (OWC) and VLC \cite{HanzoVLC}. A detailed tutorial on quantum key distribution (QKD) designed for satellite channels may be found in \cite{HanzoQ1} and in \cite{HanzoQ2}.

\subsection{Unlicensed Spectrum Access}
	Given the thirst for bandwidth, there is a need for advanced bandwidth-efficient transmission techniques. Therefore, unlicensed spectrum accessing, including both the traditional sub-6 GHz frequencies and the mmWave band, has become an active research topic of next-generation networks \cite{unli}. In the unlicensed spectrum \cite{unli}, interference mitigation becomes a particularly crucial research issue. The networks using unlicensed spectrum include licensed-assisted access (LAA) \cite{19}, IEEE 802.11ax \cite{20}, IEEE 802.11be \cite{20-1}, IEEE 802.11ad \cite{21}, and IEEE 802.11ay \cite{22}. Moreover, ultra-high-rate THz transmission is supported by the IEEE 802.15.3d standard utilizing both sub-THz and THz frequencies \cite{d1,d2,d3}.

\begin{figure}[!t]
\centering
\includegraphics[width=5in]{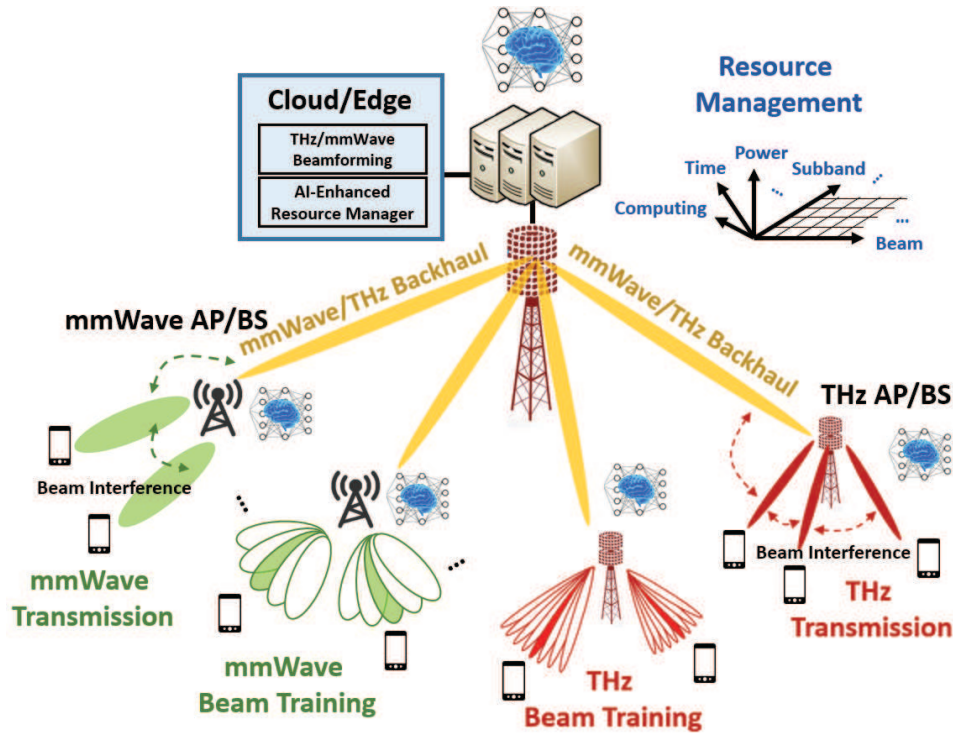}
\caption{The system architecture of coordinated THz/mmWave multiuser beamforming training.} \label{beamsim}
\end{figure}

\begin{itemize}
	\item \textbf{Sub-6 GHz Frequencies}: The goal of LAA is to deploy the legacy 4G long-term-evolution-advanced (LTE-A) system in the 5 GHz band. However, since the pathloss is increased at higher frequencies, the coverage area is reduced compared to that at 2.4 GHz. Additionally, both the IEEE 802.11ac/ax/be standard systems and weather radar systems are operated at 5 GHz, where the uplink/downlink interference control and retransmission mechanism constitute the key research issues of LAA. Therefore, LAA employs carrier aggregation (CA) in order to guarantee the target QoS of UEs. The IEEE 802.11ax system combines orthogonal frequency division multiple access (OFDMA) with sophisticated scheduling mechanisms to achieve simultaneous multiuser data transmission over separate bands \cite{20-2}. We also note that 802.11be is an enhanced version of 802.11ax, which further improves the spectral efficiency by adopting wider unlicensed bands and enhanced transmission techniques \cite{20-1}.

	\item \textbf{mmWave Frequencies}: IEEE 802.11ad/ay operates in the 60 GHz mmWave band and relies on beamforming to compensate for the high pathloss with the aid of the so-called enhanced directional multi-gigabit (EDMG) technique \cite{22-1}. However, the hidden node problem associated with the usage of 802.11ad/ay critically relying on beamforming remains a challenging unsolved issue \cite{22}. Furthermore, multiple access points (APs) have to be established for practical multiuser mmWave based EDMG transmissions relying on the 802.11ad/ay protocols. However, beamforming training is another potential challenge, which has to be tackled for finding the optimal beam direction in such complex multiuser multi-AP scenarios. Therefore, an enhanced multi-AP multiuser architecture is proposed in \cite{22-2}, which is backward compatible with the existing 802.11ad/ay protocol. To elaborate briefly, the coordination based beamforming training (CBFT) of \cite{22-2} was designed for multiple APs and multiple users with the objective of attaining near-unity successful user association ratio and a maximum tolerable beam alignment outage probability. The APs aim for flexibly tuning both the length of training frames and of the contention slots, whereas the users perform their individual association and individual beam training \cite{22-2}. As shown in \cite{22-2} quantifying both the system's latency and throughput, the CBFT imposes the lowest latency and yet achieves the highest throughput, substantially outperforming both the time division method and the conventional 802.11ad/ay protocols.

	\item \textbf{THz Frequencies}:
	The THz-band holds the promise of an abundance of bandwidths capable of fulfilling the high data rate demands of 6G. As an enhancement of the mmWave IEEE 802.11ad/ay standard, IEEE 802.15.3d is the first protocol ratified for THz transmissions \cite{d3}. It supports different applications, such as wireless backhaul/fronthaul, data centers, kiosk downloading, and even intra-chip networks \cite{d1}. The ultra-wide band proposed by the World Radio Conference 2019 (WRC-2019) spans from 2.16 GHz up to 69.12 GHz bandwidth utilization between the operating frequencies of 252.72 GHz and 321.84 GHz. However, the THz properties impose several implementation-oriented hurdles both in terms of the high-power signal-generation and signal-detection, which require carefully-crafted new THz beamforming techniques. Furthermore, due to the hostile propagation properties, 802.15.3d adopts robust low-complexity modulation schemes, such as the THz single carrier mode \textit{(THz-SC PHY)} and THz on-off keying mode \textit{(THz-OOK PHY)}, both proposed in \cite{d1,d3}. From a protocol design perspective, the network coordinator will train its THz-beams with the aid of consecutive beacons transmitted to the devices, whereas the devices will transmit directional association requests for exchanging all desired information. Given the higher beam resolution of 802.15.3d THz compared to 802.11ad/ay, it requires new low-complexity and low-overhead THz-oriented beamforming training. Furthermore, the prospective techniques of 802.15.3d conceived for the THz band include (1) simplified procedures for initial access and device discovery, (2) multiple access and interference mitigation, (3) node mobility and multiple-channel access support, and (4) up to 100 Gbps-level mid-range wireless fronthaul/backhaul capabilities using nano-antenna-arrays. To elaborate a little further, as elaborated in Fig. \ref{beamsim}, the beam network leveraging both mmWave wide-beams and THz-oriented pencil-beams becomes a promising solution in multi-spectrum accessing in a collaborative transmission manner. However, it remains an open issue regarding how to design low-complexity and low-overhead coordinative beam training and transmission mechanism under limited computing and communication resources for cross-spectrum and hybrid-radio access networks.
\end{itemize}	
	
\begin{figure}
	\centering
	\includegraphics[width=4.3in]{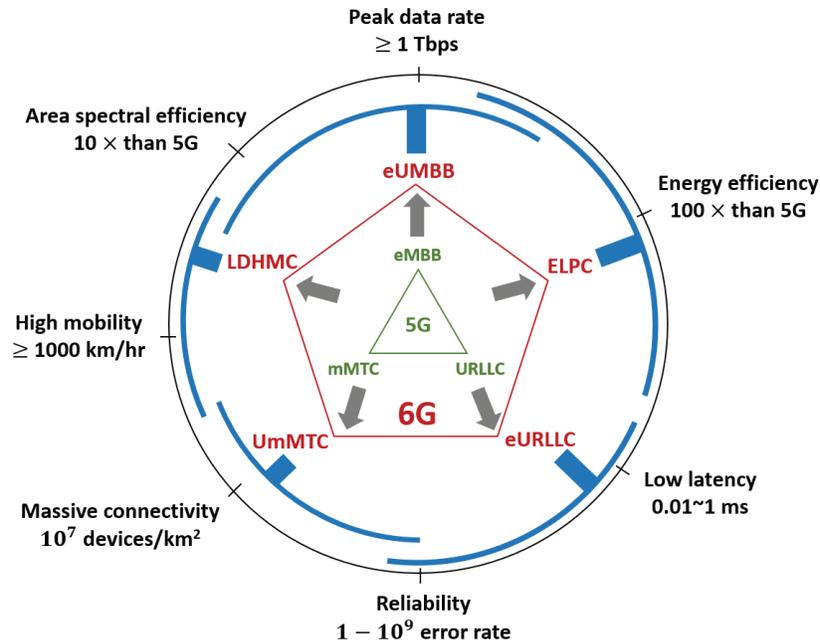}
	\caption{Multiple network services for 5G and 6G eras.} 
	\label{5G6G}
\end{figure}

\subsection{Multiple Wireless Services}
	The 5G network supports three rather different types of wireless services, including enhanced mobile broadband (eMBB), ultra-reliable and low latency communications (URLLC), as well as massive machine type communications (mMTC) \cite{24}, as depicted in Fig. \ref{5G6G}. The next-generation eMBB network services require new transmission and access technologies for achieving even higher data rates in new wireless network architectures, representing the genesis of the enhanced ultra-mobile broadband (eUMBB) philosophy. In a similar spirit, the emerging next-generation enhanced-URLLC (eURLLC) mode has beneficial applications in unmanned factories, unmanned aircraft, unmanned vehicles and intelligent transportation systems requiring instant messaging at a high reliability and low latency. Additionally, given the escalation of the number of connections, there is an increasing need for an ultra-mMTC (UmMTC) design for supporting more flexible, efficient, low-latency, highly-adaptive protocols. In the next-generation hybrid services, such as the URLLC-eMBB services \cite{25}, front-end resource allocation and hybrid numerology optimization have emerged as open issues.

\begin{figure}
	\centering
	\includegraphics[width=5in]{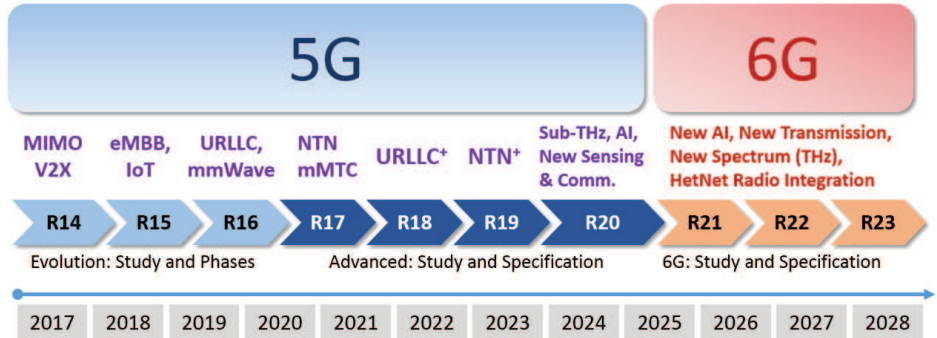}
	\caption{Tentative timeline and key technologies from 5G, advanced and 6G in 3GPP organization.} 
	\label{gpp}
\end{figure}

	As defined in the 3rd Generation Partnership Project (3GPP) specifications, the network's functional split \cite{6} is capable of supporting the flexible configuration of the entire core network and its devices for performing either centralized or distributed computing. In addition to the above-mentioned services, the research community is also discussing the conception of both long-distance and high-mobility communications (LDHMC) as well as of extremely low-power communications (ELPC) \cite{1}, as also illustrated in Fig. \ref{5G6G}. 
Moreover, as a prospect potential specification by 3GPP \cite{white1, white2}, they release the tentative timeline and key technologies from 5G and its advanced as well as 6G-era, as summarized in Fig. \ref{gpp}. Terrestrial use cases is proposed at early stage from Release 14 to 16, whereas versatile conceptions of architectures and spectrum utilization are leveraged in advanced version of 5G from Release 17 to 20. While, from Release 21, powerful AI techniques and new transmission/spectrum for community usage are tentatively conceived as 6G services in heterogeneous radios. As a result, it becomes imperative to integrate the networks both horizontally and vertically in support of high transmission rates, full coverage of remote areas, high-mobility, and lower-power IoT devices.

\section{Facet 2: Next-Generation Networking}

\subsection{Network Softwarization of SDN/NFV}
	To fulfill the challenging specifications of next-generation networks, SDN research focuses both on automatic network management and on the optimization of traffic management, which includes dynamic real-time automated network management, routing optimization, load-balancing, multi-path routing, quality management of service routes, and automatic repair of faltering routes \cite{31,32, acm2}. Network function virtualization (NFV) partitions the real network into multiple independent virtual networks, which have their individual operating resources in support of heterogeneous service qualities. The security challenges are attacks against the controller and blocking of network services \cite{33}. In a nutshell, NFV becomes an important research topic, which increases the flexibility of network deployment and reduces both the device costs and operating costs. We can further improve the system performance by combining the SDN principles with optimized management of the tele-traffic routes. Hence, NFV combined with management and orchestration (NFV-MANO) constitute pivotal research issues to be explored in the field of network softwarization \cite{34}.

\subsection{Next-Generation Packet Core Networks}
	The conventional mobile core network is constituted by specific hardware and software, including a mobility management entity (MME), as well as serving gateways (SGWs) and home subscriber server (HSS) units \cite{35}. Although the traditional core network relies on separate control and data link layers, the processing of packets between these two layers should still be performed in the switch and router simultaneously. With the rapid development of SDN and NFV, the mobile core network constitutes the natural platform for accommodating the control and data link layers, which leads to the potential research topic of the so-called virtualized evolved packet core (vEPC) \cite{36}. The 3GPP organization also proposed the 5G core (5GC) networking concept, including both standalone (SA) and non-standalone (NSA) options for flexibly adjusting the configuration of the control and user planes \cite{37, 38}. Integrating SDN and NFV techniques is capable of supporting flexible applications for telecom operators. It also contributes to the conception of the network slicing architecture \cite{38}, where a realistic network is partitioned into multiple QoS-guaranteed virtual networks. However, improving the flexibility and functionality of packet processing across different switches becomes one of the salient research issues in the SDN/NFV-enabled 6G core network (6GC) of the near future.

\subsection{Next-Generation Mobile Network Architecture and Management}
	Thanks to the introduction of SDN and NFV, the architecture and management of next-generation mobile networks exhibits a high grade of flexibility, intelligence and automation, including advanced mobile cloud and edge computing. It is a crucial task to adaptively assign computing resources to the cloud and edge. Both mobile edge computing (MEC) \cite{39,mec} and fog computing \cite{40} constitute important next-generation network architectures, which potentially lead to the reduction of service latency as well as to the improvement of both the spectral efficiency and QoS. Given the increased number of BSs forming heterogeneous networks, conventional manual control of the power allocation and BS deployment becomes infeasible. The advanced self-organized network (SON) \cite{41} concept subsumes SDN and NFV in support of self-configuration, self-optimization, self-healing and self-sustenance. The SON can also automatically execute the optimization of the key parameter settings \cite{HanzoSON}.

%-------------------------------------------------
\begin{table}
\footnotesize
\begin{center}
\caption {Comparison of NB-IoT, LoRa and SigFox}
\renewcommand{\arraystretch}{1.1}
    \begin{tabular}{l|lll}
       \specialrule{.2em}{.1em}{.1em} 
       \textbf{Feature} & \textbf{NB-IoT} & \textbf{LoRa} & \textbf{SigFox} \\ \hline\hline
       
       Alliance & 3GPP (2016) & \tabincell{l}{LoRa Alliance \\(2015)} & SigFox (2009) \\ \hline
       
       Spectrum & Licensed & Unlicensed & Unlicensed \\ \hline
       
       Frequency & \tabincell{l}{In-Band LTE, \\LTE Guard band,\\700--900 MHz} & \tabincell{l}{433, 780,\\ 868, 915 MHz} & \tabincell{l}{ 868,  902 MHz} \\ \hline
       
       Bandwidth & 180--200 kHz & 125--500 kHz & 100 Hz \\ \hline
       
       Modulation & \tabincell{l}{UL: SC-FDMA \\DL: OFDMA} & \tabincell{l}{Chirp spread\\ spectrum (CSS)} & \tabincell{l}{UWB} \\ \hline

	   Transmission & Half-duplex & Half-duplex & Half-duplex \\ \hline          
       
       Data Rate & \tabincell{l}{UL: 130 kbps \\DL: 160 kbps} & 0.25--50 kbps & 100 bps \\ \hline
       
       Output Power & \tabincell{l}{UL: 14--22 dBm\\DL: 27--30 dBm} & 14--30 dBm & 14, 20--23 dBm \\ \hline
       
       Max Range & 15 km & \tabincell{l}{Urban: 3--5 km\\ Rural: 15 km} & \tabincell{l}{Urban: 10 km\\ Rural: 50 km} \\ \hline    
       
       Connectivity & $10^5$ devs.& $2.5 \times 10^5$ devs. & $10^6$ devs.\\ \hline
       
       Battery Life & 5--10 years & 5--10 years & 5--10 years \\ \hline

	   Cost & High & Low & Medium \\ \hline
       
       Interference & Low & High & High \\ \hline
       
       Security & High & Low & Low \\ 
      \specialrule{.2em}{.1em}{.1em}

    \end{tabular} \label{iott}
\end{center}
%\end{small}
\end{table}
%------------------------------------------------	

\section{Facet 3: Internet of Things}

\subsection{IoT Access, Sensing and Data Collection}
	The IoT is a network formed by the interactions of physical objects or by the related hardware and software, gleaning information from diverse networks constituted by heterogeneous sensor devices and controllers \cite{44}. The different IoT networks require diverse types of sensors and transmission modes, as exemplified by the Internet-of-everything, the network of personal wearables \cite{45}, industrial IoT (IIoT) \cite{46,47}, intelligent home services and even the underwater Internet \cite{HanzoWater}. It is critical to design a mechanism for efficiently sensing the environments for collecting data and for uploading information to the processing server. However, there are substantial challenges, such as the coordination of different protocols for improving the system's power consumption, capacity and spectral efficiency \cite{48}.

%\begin{figure*}
%\centering
%\subfigure[]{\includegraphics[width=1.9in]{iotlora} \label{iot1}}
%\subfigure[]{\includegraphics[width=2.45in]{Outdoor1_drop} \label{iot2}}
%\subfigure[]{\includegraphics[width=2.45in]{Outdoor1_th} \label{iot3}}
%\caption{Experimental environments and corresponding performance results in \cite{49} comprising $5$ normal LoRa-based IoT devices and $2$ IoT devices with emergency. (a) Experimental setting of LoRa-based IoT gateway and sensor IoT device, (b) packet drop rate and (c) system throughput.} \label{iotsim}
%\end{figure*}	
	
	The versatile features of IoT transmission are capable of facilitating diverse configurations in support of either large-scale access, or long transmission distances, low-power and/or low-rate operation at low deployment cost \cite{48}. In addition to machine-to-machine (M2M) type communications, the family of advanced IoT-based protocols include Zigbee, ZWave, Bluetooth Low Energy (BLE), Bluetooth 5.0, SigFox, IEEE 802.11p for vehicular communications, the Long Range (LoRa) protocol \cite{49}, 3GPP Narrow Band IoT (NB-IoT), and WiFi HaLow for IEEE 802.11ah, which are so-called low-power wide area network (LPWAN) solutions \cite{iot,50}. The comparison of popular IoT technologies including NB-IoT, LoRa and SigFox is summarized in Table \ref{iott}. To elaborate a little further, NB-IoT \cite{51} is the standardized protocol relying on 5G BSs for providing convenient IoT access based on the existing infrastructure. Furthermore, wireless wide area networks based on LoRa \cite{50} and IEEE 802.11ah are capable of offering private or public network deployment for industrial applications. 
	
\begin{figure}
\centering
\includegraphics[width=4in]{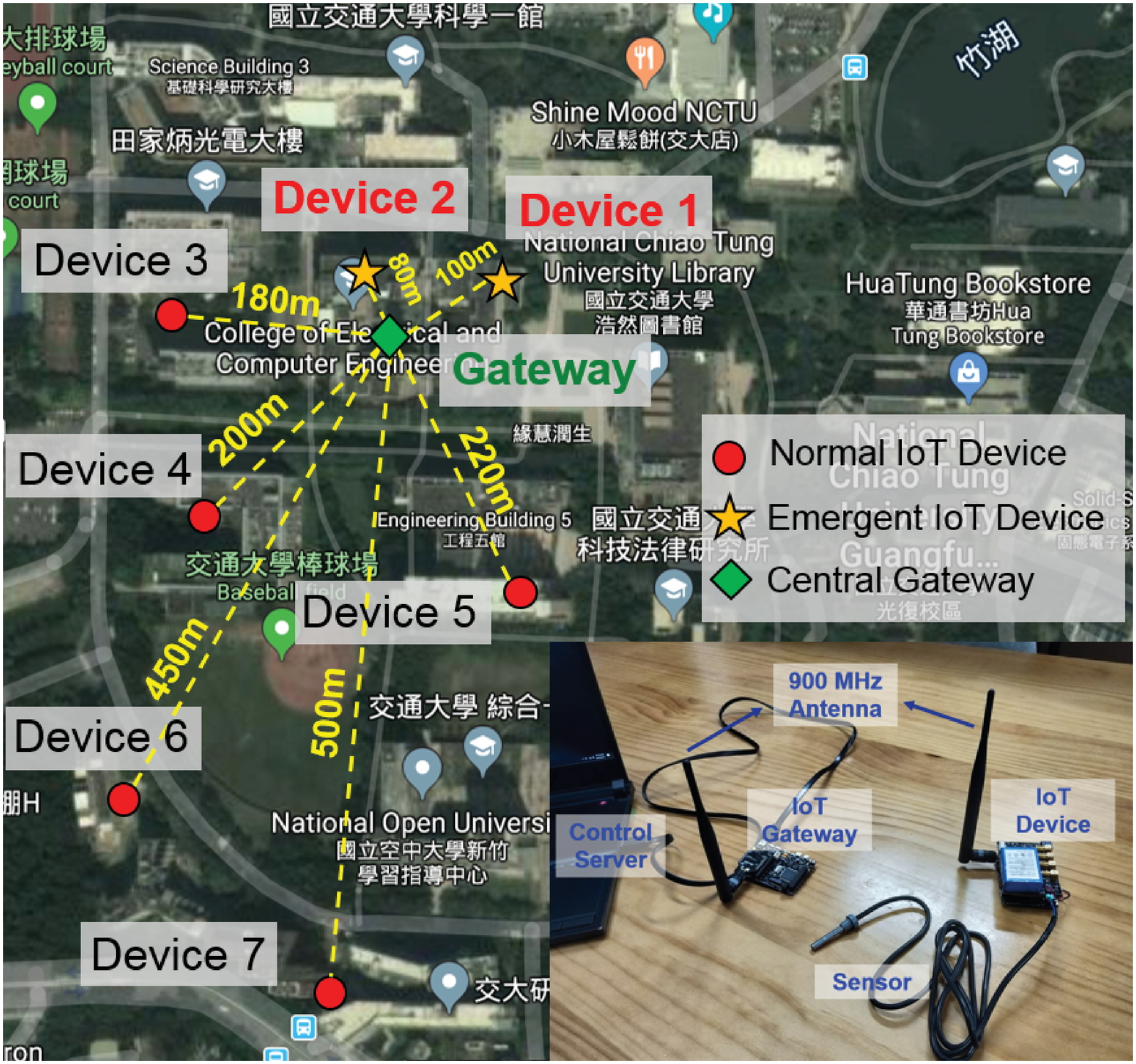}
\caption{Experimental environment of an LoRa-based IoT network \cite{49}.} \label{iotsim}
\end{figure}

	However, these revolutionary techniques should consider the potentially conflicting requirements of high area spectral efficiency, low interference, infrequent handovers, and power conservation \cite{acm5}. A range of further challenging problems are associated with data encryption \cite{52}, traffic management and resource allocation in low-power IoT networks \cite{49}. The above-mentioned IoT technologies mainly rely on using narrow band communications. Therefore, we should design advanced schemes for striking a tradeoff among the requirements of narrow bandwidth, low power consumption, tight synchronization, and limited processing complexity. The authors of \cite{49} have established an LoRa-based IoT network for the complex scenario seen in Fig. \ref{iotsim} by proposing a joint traffic-aware channel and contention backoff window size allocation (TCBA) scheme capable of handling diverse IoT traffic loads characterized by their packet arrival rates. The IoT covers multiple research areas, which requires the consideration of the overall network architecture and various heterogeneous technologies, while tackling the technical challenges of efficiency, reliability, and integration in the sensing layer, the network transport layer, the operations and management layer, as well as the application layer \cite{53,54}.

\subsection{Vehicular Networks}
As a special use case of IoT, there is an ever-increasing demand for high-performance Internet of Vehicles services \cite{v2x-1}, which stimulates substantial research. One of the most important challenges is that of connected autonomous vehicles (CAVs) relying on joint sensing, control, as well as communications \cite{v2x-7}. The associated IoT devices and sensors are carried by the vehicles with objective of detecting and tracking objects, such as pedestrians, other vehicles, and traffic signs. Moreover, it is vitally important to keep track of the status of road traffic. Furthermore, vehicle-to-everything (V2X) \cite{v2x-8} supports the wireless exchange of information between vehicles and other connected devices. For example, V2X communication is capable of supporting the IoT sensors by providing long-range detection of hazards, traffic conditions, and blind spots outside the vehicular field of view (FoV). By relying on IoT services, V2X technology improves road safety, traffic efficiency, and energy efficiency through the employment of road side units. More specific service types include vehicle-to-infrastructure (V2I), vehicle-to-vehicle (V2V), vehicle-to-network (V2N), vehicle-to-pedestrian (V2P) and vehicle-to-device (V2D) solutions relaying either on cellular-based or wireless local area network (WLAN)-based systems. The cellular V2X (C-V2X) and NR-V2X harness existing cellular NR-based networks and V2X protocols \cite{v2x-3, v2x-8}. Furthermore, as specified by the IEEE 802.11p standard, dedicated short-range communications (DSRC), which is the first V2X communication service relies on WLAN technology and supports directly both V2V and V2I services by forming vehicular ad-hoc network \cite{v2x-9}. The open challenges in IoT-V2X include their wireless channel characteristics, resource management, their heterogeneous interfaces, their dynamic topology, efficient routing and trajectory design, congestion policy, security and reliability, as well as joint optimization of sensing, control and communications \cite{v2x-2, v2x-4, v2x-5, v2x-6, v2x-10}.

\subsection{Social IoT Network}

	Social IoT (S-IoT) networks are virtual social networks formed by a group of IoT devices belonging to people having similar interests \cite{56-1}. The S-IoT integrates IoT networks, which rely on proximity services (ProxSe) \cite{56-2}. Note that under a ProxSe scenario members of mobile community networks must be geographically adjacent and are capable of directly accessing D2D or vehicular communications. The S-IoT can also be regarded as a human social network supporting efficient services or facilitating the interaction of sensor devices. The S-IoT is also capable of reusing social networking modules for IoT networks, including pedestrian/vehicular mobile social networks \cite{56-3}. The potential future applications will include advertising, geographic data or content sharing, social networking platforms, robotic systems, gaming platforms, the relaying of data from the users or IoT devices, including the enhanced driving safety, roadside information access, unmanned aerial vehicles, autonomous driving, and environmentally-friendly vehicles.

	In this context, \textit{network science} aims for analyzing the resultant complex networks in terms of their topology, dynamic characteristics, behaviors, functions and diverse attributes by relying on graph theory \cite{social1}, social network theory \cite{HanzoSocial}, statistical physics \cite{social2}, biology \cite{social3}, and social science \cite{social4}. Some of the open research issues in 6G include but are not limited to: (1) the network's topology, connectivity, resilience and robustness; (2) information dissemination assisted by the epidemic network model and network inference \cite{56-4}; (3) the conception of advanced analytic tools relying on graph theory \cite{56-5} and game theory \cite{56-6}.
		
\subsection{Security and Privacy}
	In recent years, numerous network attacks have taken place threatening user security and privacy \cite{58,secure}. Hence, rapid advances took place in privacy enhancement as well as in physical layer security (PLS) \cite{HanzoPLS}. However, the existing security and privacy protection techniques only consider a single functionality and service, which has to be extended to multiple services \cite{60}. Furthermore, next-generation quantum cryptography and post-quantum cryptography have numerous challenging open problems \cite{61}. 

	The privacy issues of IoT applications in wearable device networks, smart grids, and vehicular networks are in their infancy. For example, under wireless body area networks (WBANs) relying on the IEEE 802.15.6 standard \cite{65}, wearable device protocols have to be redesigned for ensuring the security and privacy of wearable devices. The privacy-aware strategies of smart grids associated with electricity prediction and billing also require substantial future research \cite{66}. Moreover, the security-critical issues of vehicular networks also require further advances, with special attention on emergency messages, authentication, and ultra-low latency communications between road side units (RSUs) and vehicles in the face of high mobility. Additionally, the security and privacy of intelligent cloud and edge networks also constitute important research topics \cite{68}.

\section{Facet 4: Wireless Positioning and Sensing}

\subsection{Outdoor Positioning}
	The most widely used outdoor positioning \cite{69} system at the time of writing is the global positioning system (GPS), which has a limited position accuracy and limited coverage owing to its high signal loss. Its shortcomings might be mitigated by beneficially harnessing the signals gleaned from cellular BSs \cite{70}, for example, for the location tracking of vehicles and UAVs \cite{71, ourcsiKI}. However, the excessive Doppler shift of high-speed movement constitutes a critical challenge in the physical layer of high-accuracy tracking \cite{72}. By relying on sophisticated AI techniques \cite{73}, we can design compelling outdoor applications around regional points of interests \cite{74}. The associated space-time based positioning information can be beneficially exploited for precisely tracking specific user trajectories. Furthermore, a whole suite of challenging but promising services may be conceived upon integrating long-distance and/or low-power IoT networks into the existing cellular vehicle tracking, for example.	

\subsection{Indoor Positioning}
	Since the existing GPS system has almost no indoor coverage, the indoor positioning systems typically rely on WiFi signal strength measurements and pre-recorded radio frequency (RF) maps termed as fingerprinting \cite{indoor,75}. However, the precision of fingerprint-based indoor positioning techniques critically hinges on the stability of wireless signals and on the establishment of large databases, which is extremely laborious \cite{76,77}. This task may nonetheless be mitigated with the aid of spatial skeleton databases inferred from indoor map information \cite{ourcsiJack}. As a design alternative, lower-power Bluetooth \cite{ourcsiblue} and ultra-wide band (UWB) scenarios may also be adopted by the indoor positioning systems. In order to improve the attainable positioning accuracy, we can deploy various micro sensors such as RF identification (RFID), infrared, ultrasonic, visible light \cite{78} as well as other smart devices, including accelerometers, gyroscopes, magnetometers, air pressure, ambient sound and laser sensors. 
	
\begin{figure}
	\centering
	\includegraphics[width=4.5in]{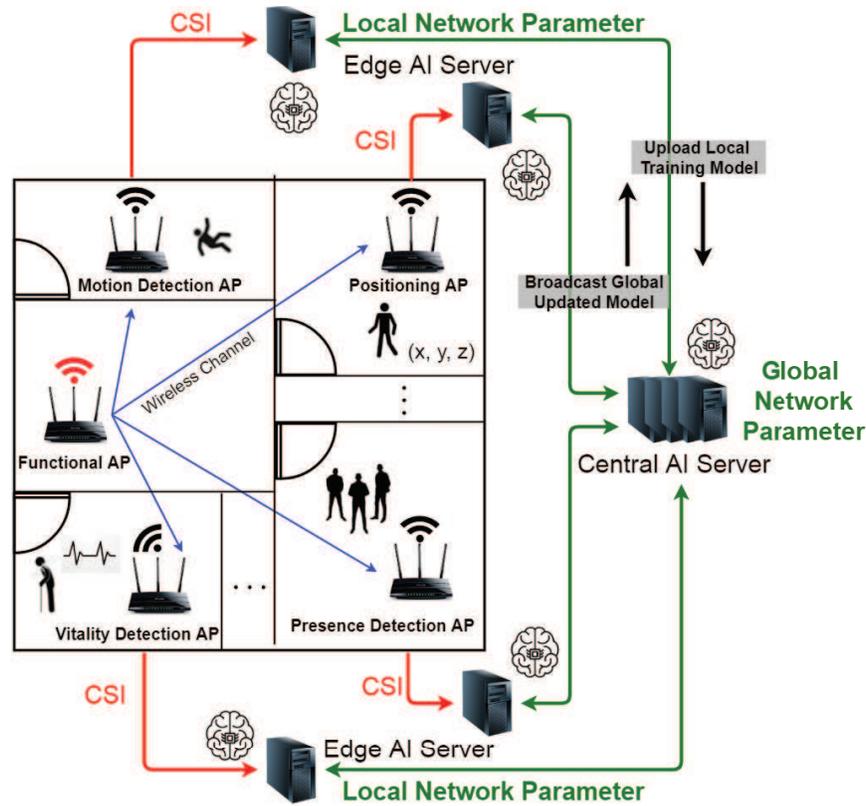}
	\caption{Device-free wireless positioning and sensing detection with the deployed functional APs, edge and central AI servers. The CSI dataset is collected via wireless channels between AP pairs. Local network parameters are learned by edge AI servers, whereas global model updating and broadcasting are performed by a central AI server.} 
	\label{sen}
\end{figure}		
	
	Moreover, when aiming for centimeter-level positioning accuracy, the channel state information (CSI) has also been widely adopted to collect positioning data \cite{79,80, ourcsichu, ourcsiKI} in order to glean a more complete signal profile of frequency, power, and latency than simply harnessing the received signal strength \cite{indoor}. Recent studies have also been conducted in the mmWave \cite{81, ourcsimmw} and THz bands, while relying on beamforming techniques \cite{82} to collect higher-dimensional signal sources for wireless positioning and sensing. In a nutshell, the positioning receiver of the above-mentioned so-called device-based regimes is required to collect measurement data concerning known reference points when performing fingerprint-based indoor localization.

	However, the evolutionary trend is to design device-free indoor positioning \cite{83,ourcsi2, ourcsichu, ourcsiKI} algorithms operating without the aid of wearable devices. The corresponding overall architecture is illustrated in Fig. \ref{sen}, which is established by relying on multiple APs associated with AI-assisted edge and central servers. Whilst Fig. \ref{sen} might seem complex, it provides an easy-reading anecdotal portrayal of the whole gamut of ideas under discussion by the scientific community. As the 6G standardization evolves further towards a broad global consensus, this figure may be reconstructed according to the harmonious confluence of ideas elaborated as follows. As shown in Fig. \ref{sen}. The databases can be gleaned from the front-end AP by measuring and scrutinising the specific fluctuation of the received signals, which characterize the particular nature of indoor activities, including motion, positioning, presence and vitality detection. These issues will be further detailed in the following subsection in the context of Fig. \ref{csisim}. The edge server carries out the local training of the network's model parameters related to its corresponding behavior, and the resultant trained local models will be merged into a global model by the central server. This device-free positioning philosophy is especially suitable for application scenarios where no wearable devices are available, including continuous tracking and ushering \cite{newtrack}, which allows us to simply monitor the wireless signals without revealing any user identities \cite{84}. Furthermore, in hostile indoor scenarios of oil tankers, mining pits or complex plants \cite{85}, it becomes extremely challenging to glean accurate positioning information. Therefore, the key research issues of indoor positioning include object tracking, trajectory modelling and the associated parameter optimization of device-free positioning.

\subsection{Wireless Indoor Detection}
	Given the rapid development of device-free indoor positioning \cite{83, ourcsi2, ourcsichu, ourcsiKI}, the application scenarios of smart homes, green buildings, factory manufacturing and healthcare will all substantially benefit. Moreover, these fine-grained detection techniques will also find innovative applications in pedestrian path tracking, presence detection, motion detection, and vitality detection \cite{detect}.
	
\begin{itemize}
	\item \textit{Pedestrian Tracking} aims for tracking a human's walking trajectory within a specific area \cite{detect1}. We can distinguish their potential behaviors through historical data. However, due to the complexity of overlapped signals in the time and frequency domains, it is quite challenging to carry out multi-object tracking, which remains an open research issue.
	
	\item \textit{Presence Detection} infers the existence of people in indoor environments based on the variation of received signals \cite{detect2, detect2-1}, but the challenge is that not only human presence, but a range of other events may result in time-varying signals. False detection may take place even owing to humidity and temperature changes in the air, and due to the unpredictable locations of interfering objects, which requires time-consuming replenishment of the database. To elaborate a little further, presence detection across different rooms \cite{ourcsichu, ourcsiKI} imposes substantial challenges on the associated signal analysis, where the associated attenuation as well as multipath effects, which should be jointly taken into consideration. As the detection coverage area is expanded, it becomes imperative to strike a balance among deployment costs, implementation complexity and detection accuracy.
	
	\item \textit{Motion Detection} aims for detecting human behavior and movements such as standing, hand-waving, falling, slow walking, jumping and so on \cite{detect3, detect3-1, ourcsihand}. Different movements potentially lead to distinct signal changes in the wireless paths. Hence, advanced algorithms may be conceived for detecting different human behaviors with reasonable accuracy. However, the network topology should be carefully designed to avoid interferences from other objects, which may severely deteriorate the accuracy of motion detection.
	
	\item \textit{Vitality Detection} analyzes wireless signals for detecting slight human feature changes, e.g., breathing rate \cite{detect4, detect4-1} and heart rate changes. With the aid of breath/heart rate estimation and prediction, a carer of the elderly may be notified if abnormal heartbeat, arrhythmia, apnea as well as severe snoring occurs. Due to more subtle changes in the human body compared to the surroundings, existing advances are confined to a small area or very short distances in order to provide adequate detection accuracy.
\end{itemize}
	
\begin{figure*}
\centering
\subfigure[]{\includegraphics[width=3in]{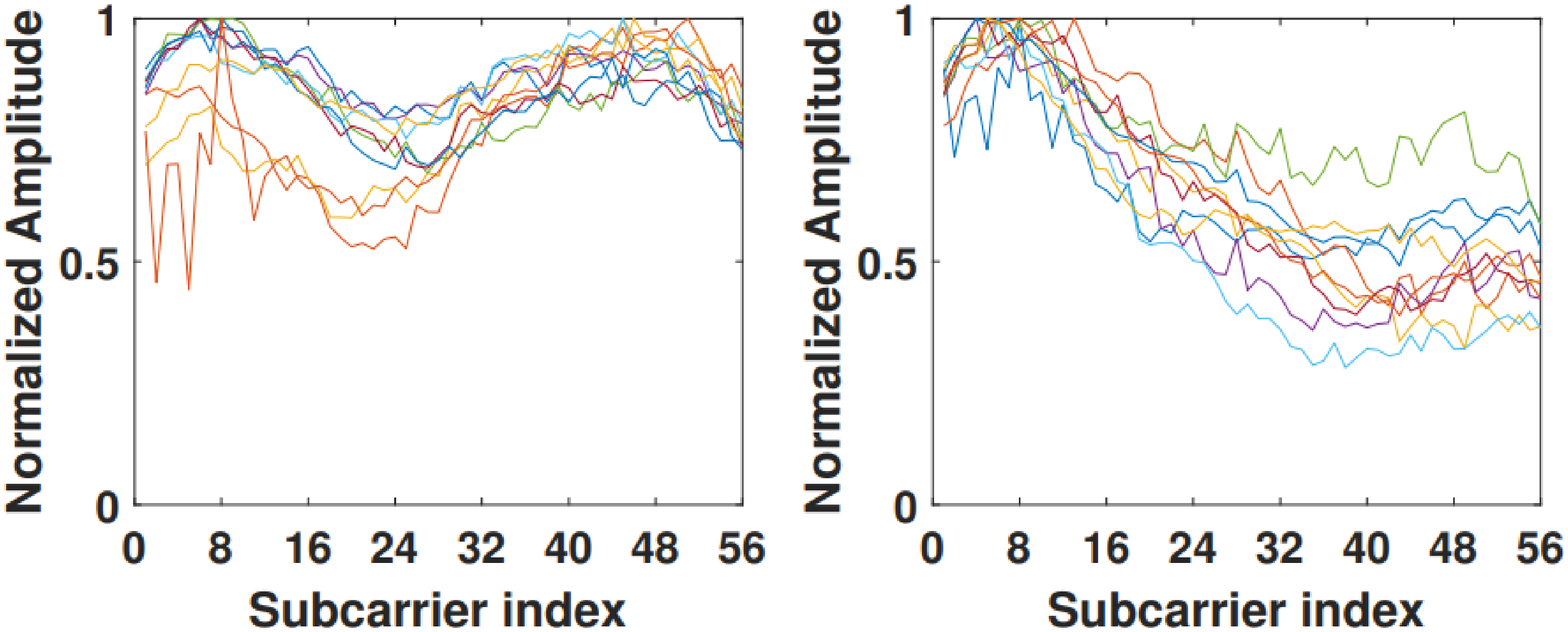} \label{csi1}}
\subfigure[]{\includegraphics[width=3in]{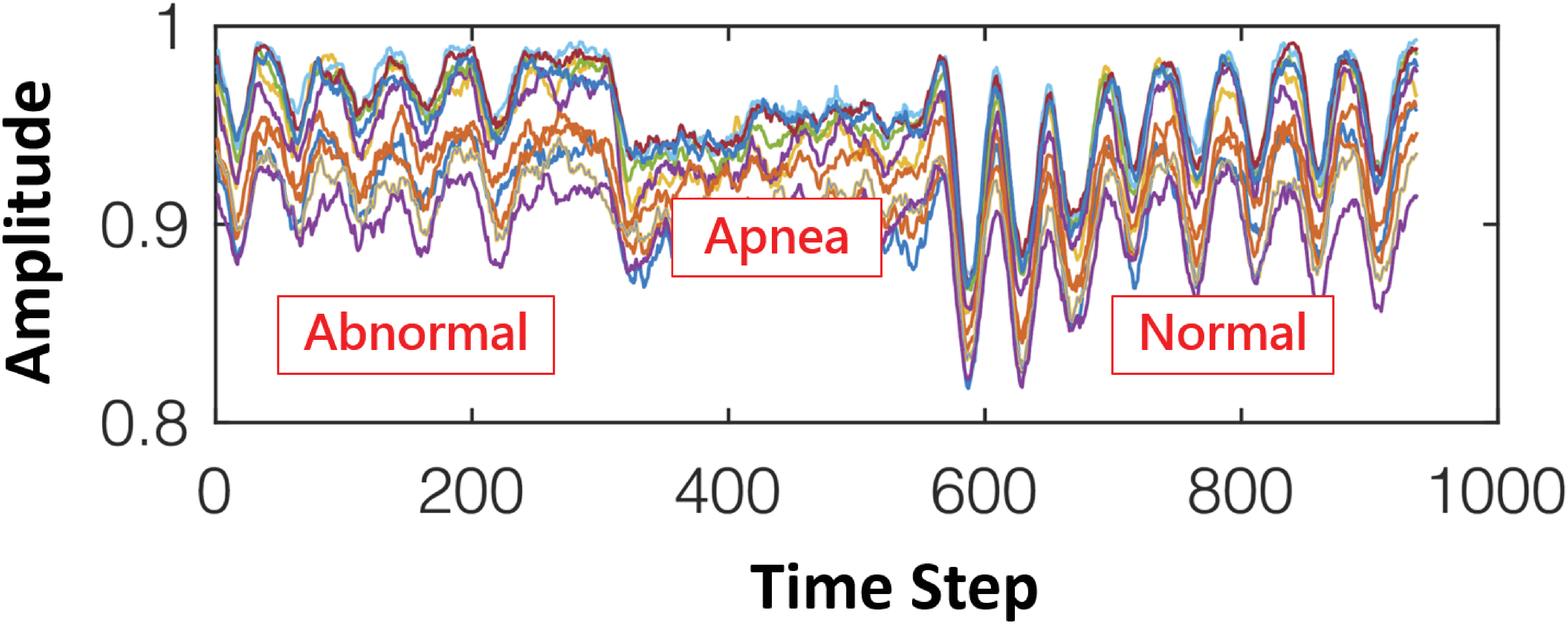} \label{csi2}}\\
\subfigure[]{\includegraphics[width=2in]{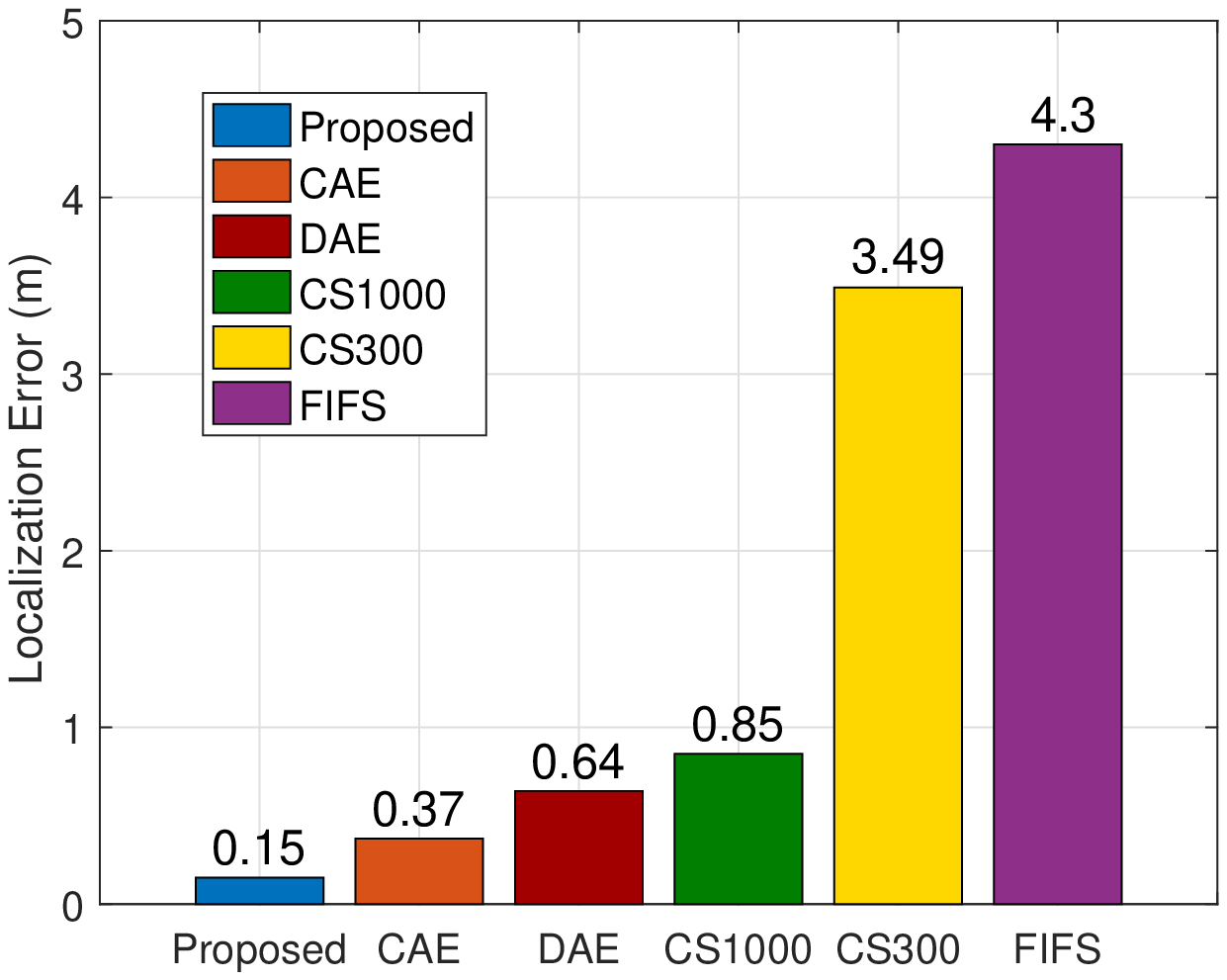} \label{csi3}}
\subfigure[]{\includegraphics[width=2in]{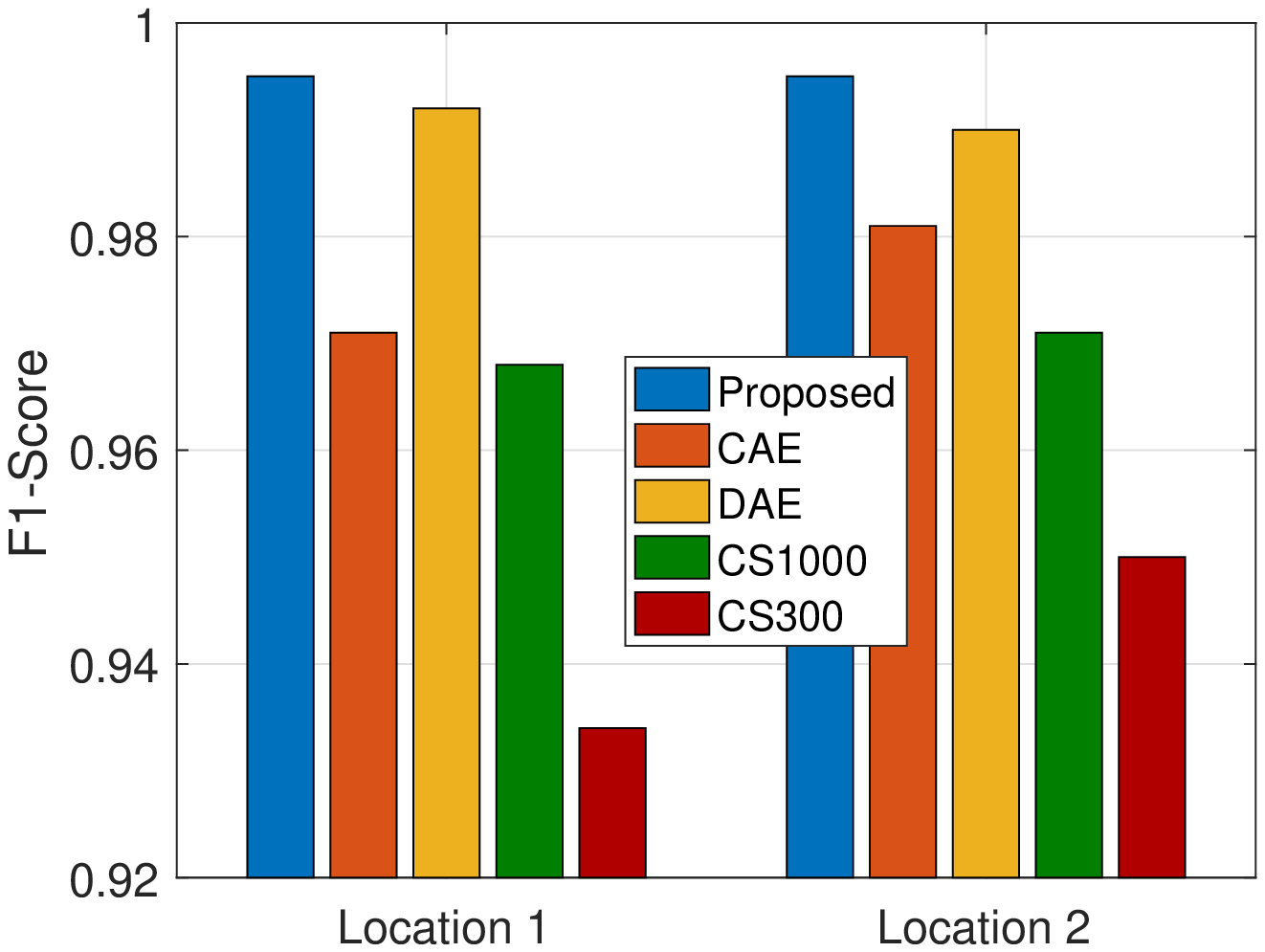} \label{csi4}}
\subfigure[]{\includegraphics[width=2in]{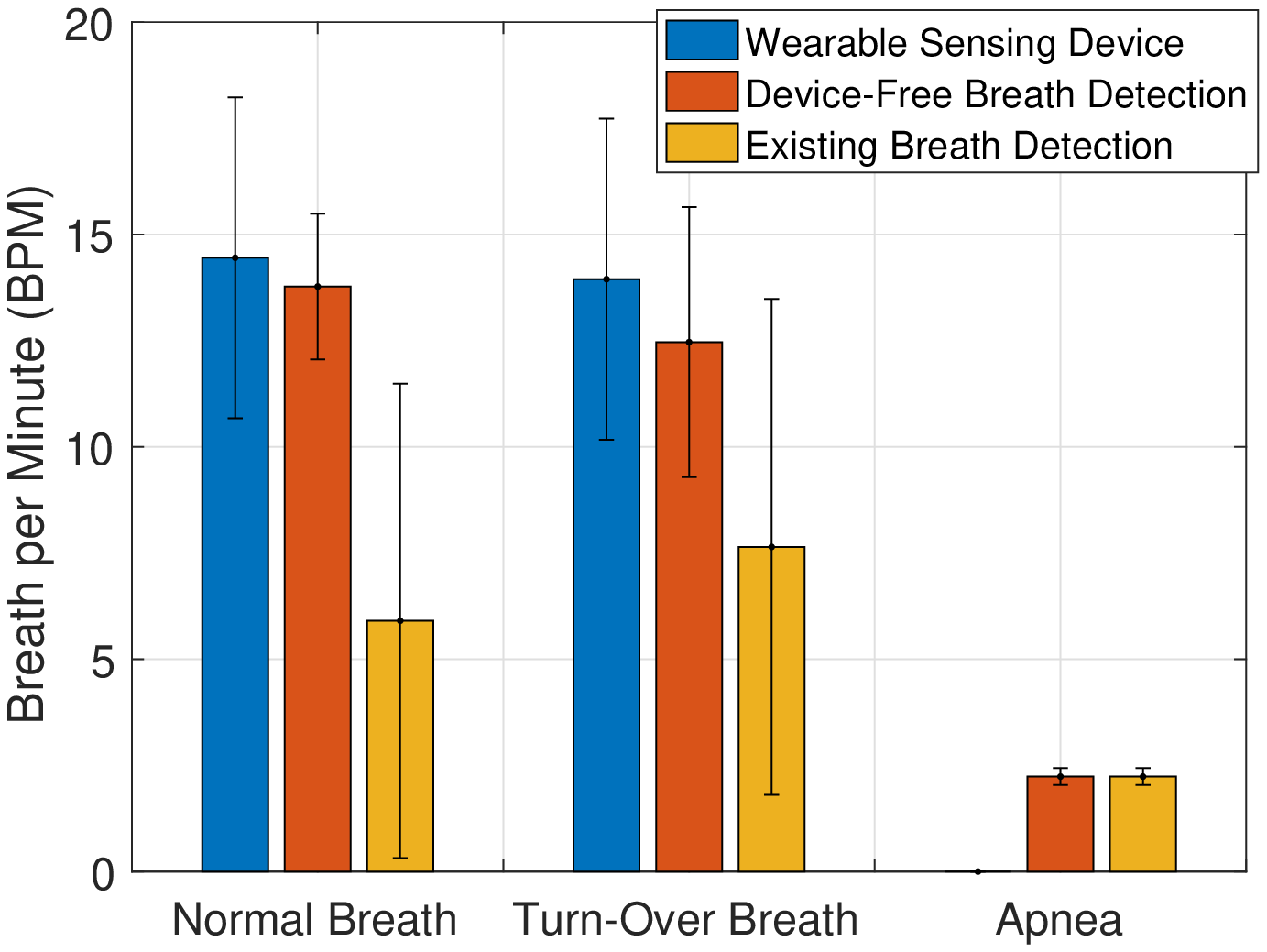} \label{csi5}}
\caption{Experimental results of (a) CSI for presence observations and (b) CSI for abnormal/normal/apnea breathing scenarios and the corresponding performances of (c) localization \cite{ourcsi1}, (d) presence detection \cite{ourcsi2} and (e) breath rate detection in terms of localization errors, F1-score and rate of breath per minute (BPM), respectively.} \label{csisim}
\end{figure*}	

Most existing techniques utilize wearable devices for maintaining detection accuracy; however, wearing devices may lead to potential inconvenience and frequent battery recharging \cite{HanzoLifeTime}. It becomes imperative to enhance the signal processing techniques in the physical layer for improving the attainable device-free detection precision at a reduced computational complexity. Regarding the techniques illustrated in Fig. \ref{sen}, the authors of \cite{ourcsi1, ourcsi2} have established an AI-empowered device-free WiFi-based CSI learning platform for both positioning as well as for presence and vitality detection. The associated experimental results are depicted in Fig. \ref{csisim}. It can be readily observed in Fig. \ref{csi1} that the CSI difference indicates the presence and absence of humans. By contrast, Fig. \ref{csi2} shows three breathing scenarios, including abnormal, apnea, and normal states. In comparison to the existing positioning methods found in the open literature and characterized in Fig. \ref{csi3}, the device-free machine learning based CSI positioning scheme proposed in \cite{ourcsi1} achieves the lowest localization error of $0.15$ meter. Moreover, the CSI-based presence detection framework proposed in \cite{ourcsi2} and characterized in Fig. \ref{csi4} has the highest detection accuracy in terms of the F1 metric defined in \cite{ourcsi2}. In Fig. \ref{csi5}, it is observed for vitality detection that the device-free breath detection may approach the performance of wearable sensing devices.

%-------------------------------------------------
\begin{table*}[!ht]
\scriptsize
\begin{center}
\caption {AI for 6G Communication and Networking Technologies}
\renewcommand{\arraystretch}{1.2}
    \begin{tabular}{|l|l|l||l|}
       \specialrule{.2em}{.1em}{.1em}
       \tabincell{l}{AI Schemes}  &  \tabincell{l}{Machine \\ Learning} & \tabincell{l}{Deep \\ Learning} & \tabincell{l}{Potential Applications and Solutions}\\ \hline
       %%%%%%%%%%%%%%%%%%%%%%%%%%%%%%%%%%
       \tabincell{l}{Supervised \\ Learning}
       & \tabincell{l}{
       $\checkmark$ \ SVM \\ 
       $\checkmark$ \ Bayesian  \\ 
       $\checkmark$ \ KNN \\ 
       $\checkmark$ \ LDA \\ 
       $\checkmark$ \ Decision Tree \\} 
       
       & \tabincell{l}{
       $\checkmark$ \ DNN \\ 
       $\checkmark$ \ CNN \\
       $\checkmark$ \ RNN \\
       $\checkmark$ \ LSTM \\ 
       $\checkmark$ \ GNN} 
       
       & \tabincell{l}{
       $-$ Optimum UM-MIMO beamforming for mmWave/THz (DNN, CNN)\\
       $-$ Channel and traffic classification and prediction (SVM, LDA, Bayesian) \\
       $-$ Device-free positioning and detection sensing (KNN, RNN, LSTM)\\
       $-$ Automatic management of integrated 6G RAT (GNN, Decision Tree)\\
       $-$ Mobility-based handover mechanism (SVM, LDA LSTM)\\
       $-$ Big data processing for IoT (DNN, CNN, RNN, LSTM)\\
       $-$ 6G multi-network and multi-service optimization (DNN, GNN)\\
       } \\ \hline
       
       %%%%%%%%%%%%%%%%%%%%%%%%%%%%%%%%%%
       \tabincell{l}{Unsupervised \\ Learning}
       & \tabincell{l}{
       $\checkmark$ \ K-means \\ 
       $\checkmark$ \ PCA \\
       $\checkmark$ \ SVD \\ 
       $\checkmark$ \ HMM \\ 
       $\checkmark$ \ EM \\ } 
       
       & \tabincell{l}{
       $\checkmark$ \ Autoencoder \\ 
       $\checkmark$ \ GAN } 
       
       & \tabincell{l}{
       $-$ 6G network data augmentation (GAN)\\
       $-$ Wireless channel detection and generation (SVD, Autoencoder, GAN) \\
       $-$ Grant-free transmissions (HMM, EM)\\
       $-$ Server data dimension reduction (PCA)\\
       $-$ High-precision trajectory tracking (K-means, HMM)\\
       $-$ 6G Network function and BS deployment (K-means)\\
       $-$ Security and privacy enhancement (Autoencoder, GAN)\\
       $-$ Unlicensed spectrum sensing (HMM, EM)\\
       
       } \\ \hline
       
       %%%%%%%%%%%%%%%%%%%%%%%%%%%%%%%%%%
       \tabincell{l}{Reinforcement \\ Learning}
       & \tabincell{l}{
       $\checkmark$ \ Q Learning \\ 
       $\checkmark$ \ Monte Carlo } 
       
       & \tabincell{l}{
       $\checkmark$ \ DQN \\ 
       $\checkmark$ \ DDPG \\
       $\checkmark$ \ IRL} 
       
       & \tabincell{l}{
       $-$ Cloud/edge computing resource management (Q learning, DQN, DDPG)\\
       $-$ Enhanced 6G RRM for diverse tele-traffic demands (DQN, DDPG, Monte Carlo)\\
       $-$ Network numerology adaptation (Q learning, DQN, DDPG)\\
       $-$ New user contention and accessing schemes (Q learning)\\
       $-$ QoS-guaranteed virtual networks (DQN, DDPG)\\
       $-$ Transmission and traffic scheduling (Q learning, DQN, DDPG, Monte Carlo) \\
       $-$ Network power control based on wireless experts' experience (IRL, DDPG)
       } \\ \hline
       
       %%%%%%%%%%%%%%%%%%%%%%%%%%%%%%%%%%
       \tabincell{l}{Distributed \\ Learning}       
       &  - &  - 
       & \tabincell{l}{
       $-$ Parallel computing in 6G core \\
       $-$ Multi-task oriented communications and networking \\} \\ \hline
       
       \tabincell{l}{Federated \\ Learning} 
       & - & - 
       & \tabincell{l}{
       $-$ Security-/privacy-aware strategies for multiusers and multi-services \\ 
       $-$ Coordination and cooperation among different 6G networks} \\ \hline
       
       \tabincell{l}{Transfer \\ Learning}
       & - & - 
       & \tabincell{l}{ 
       $-$ Rapid model establishment for SDN/NFV based SON \\
       $-$ Positioning and detection in different coverage areas} \\ 
       \specialrule{.2em}{.1em}{.1em}

    	\multicolumn{4}{l}{ \footnotesize \tabincell{l}{SVM: support vector machine, KNN: k nearest neighbor, LDA: linear discriminant analysis, DNN: deep neural network, \\ CNN: convolutional neural network, RNN: recurrent neural network, LSTM: long short-term memory, \\GNN: graph neural network, PCA: principal components analysis, SVD: singular value decomposition, \\HMM: hidden Markov model, EM: expectation-maximization, GAN: generative adversarial network, \\DQN: deep Q network, DDPG: deep deterministic policy gradient, IRL: inverse reinforcement learning }}
    	
    \end{tabular} \label{aitable}
\end{center}
%\end{small}
\end{table*}
%------------------------------------------------

\begin{figure*}
\centering
\subfigure[]{\includegraphics[width=2in]{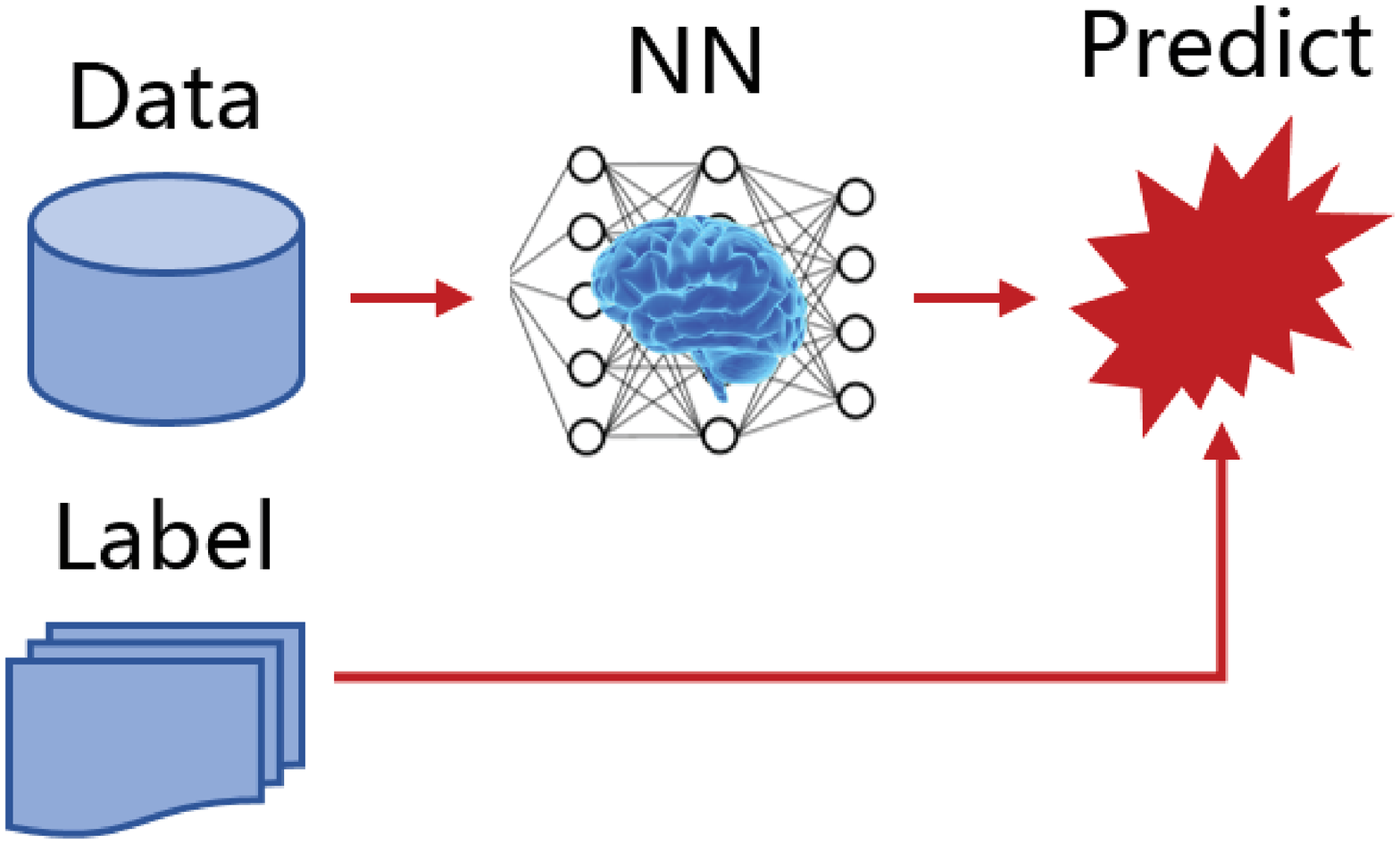} \label{aiplot1}}
\subfigure[]{\includegraphics[width=2in]{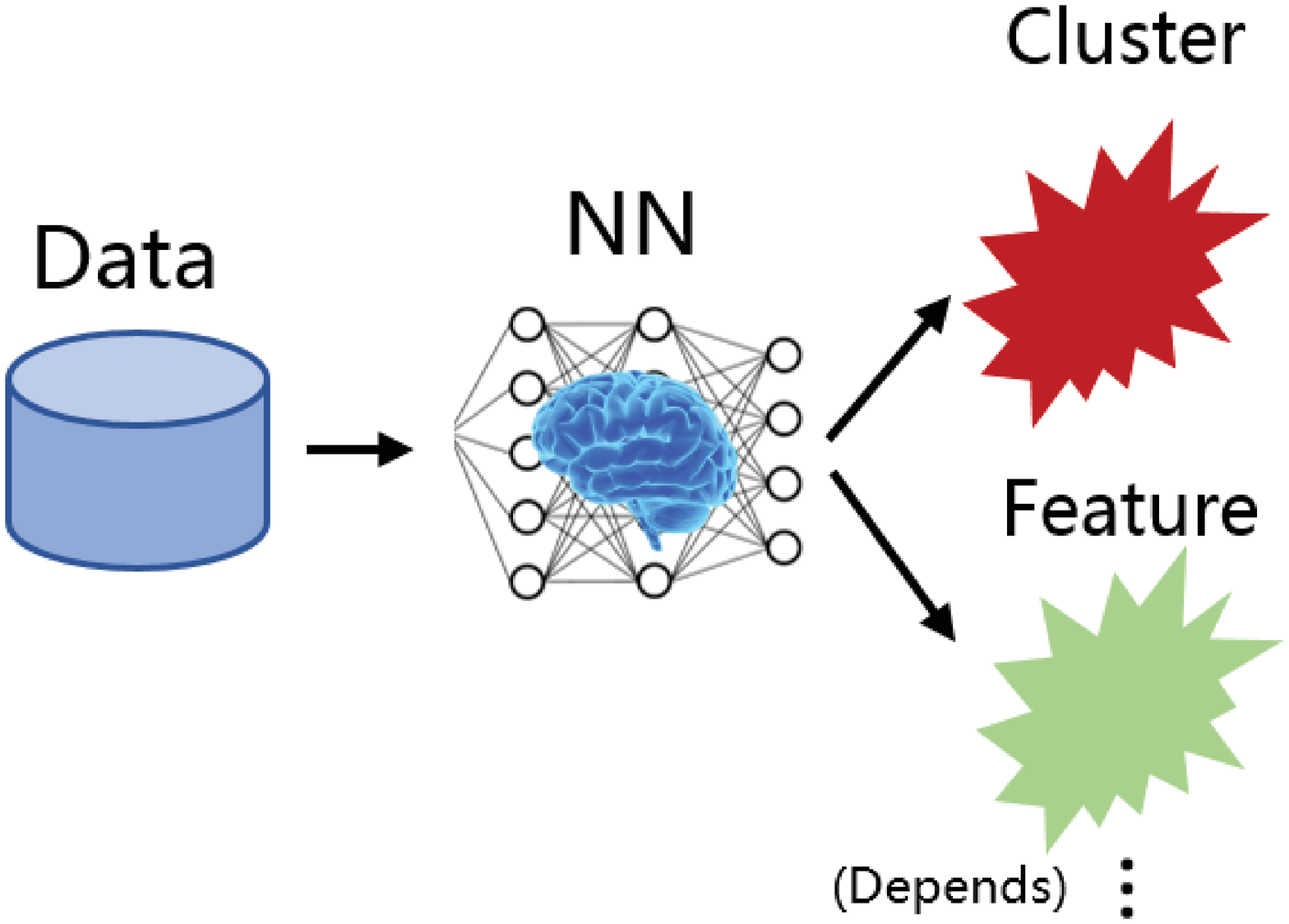} \label{aiplot2}}
\subfigure[]{\includegraphics[width=2in]{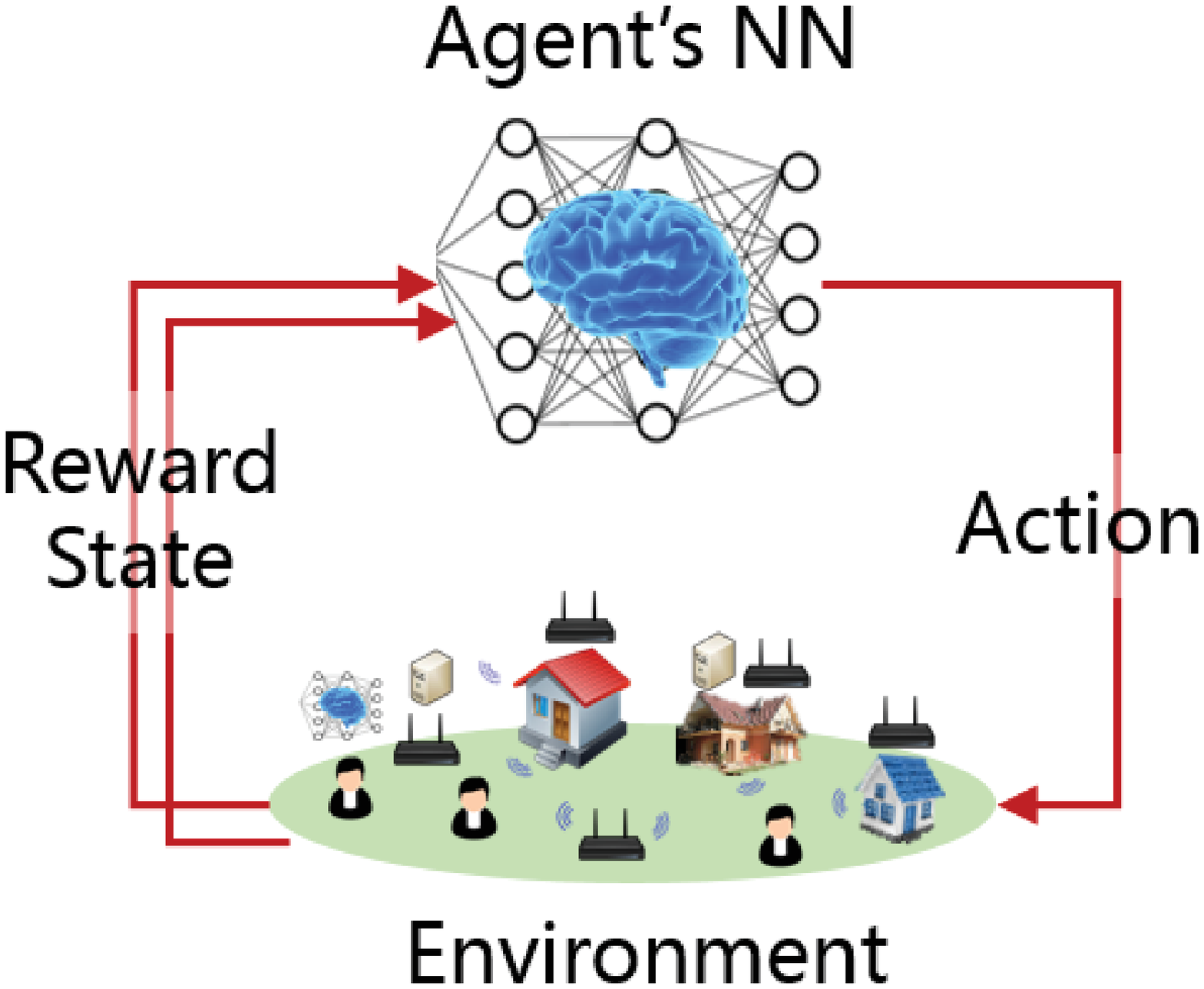} \label{aiplot3}}
\subfigure[]{\includegraphics[width=2in]{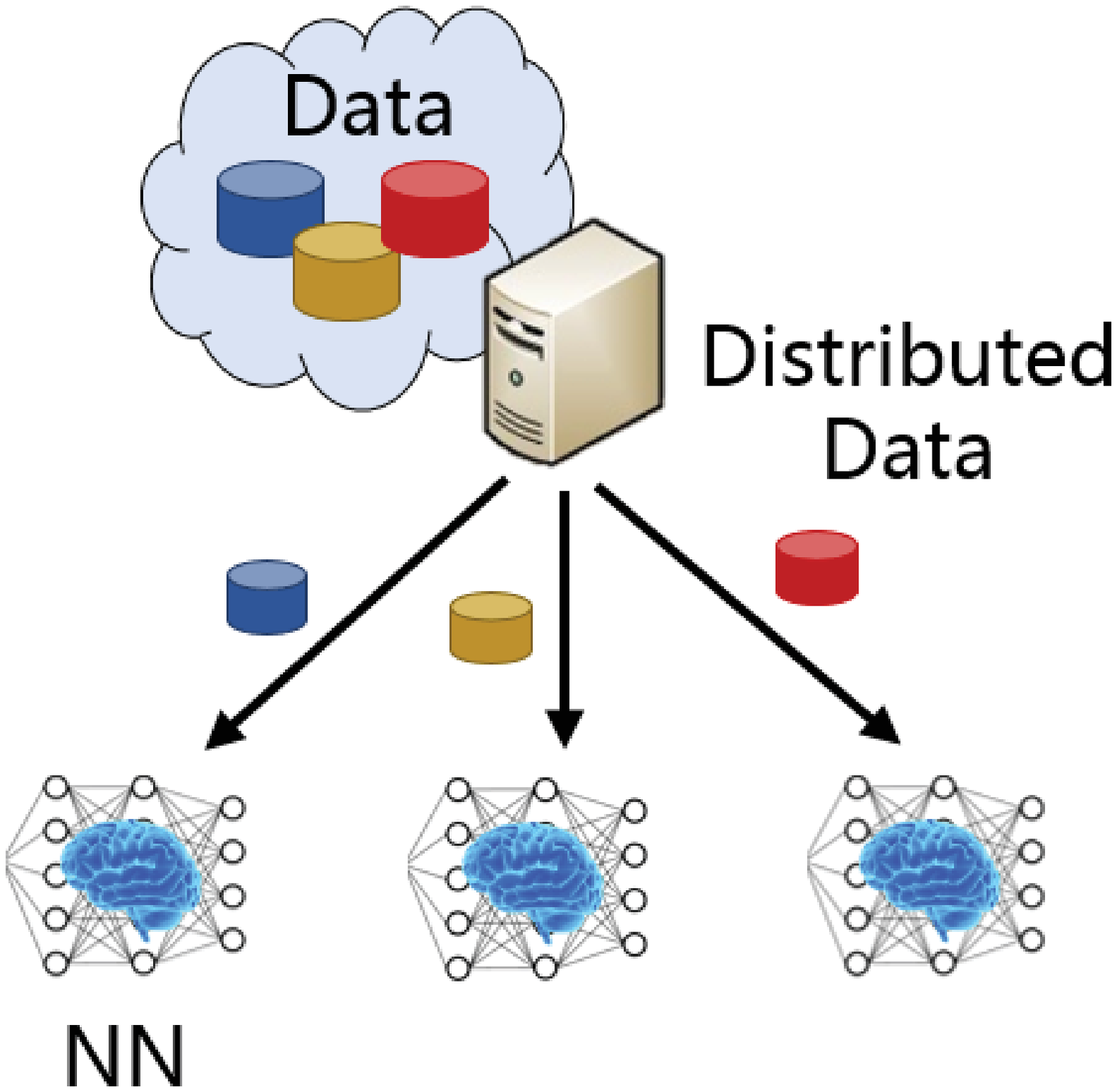} \label{aiplot4}}
\subfigure[]{\includegraphics[width=2in]{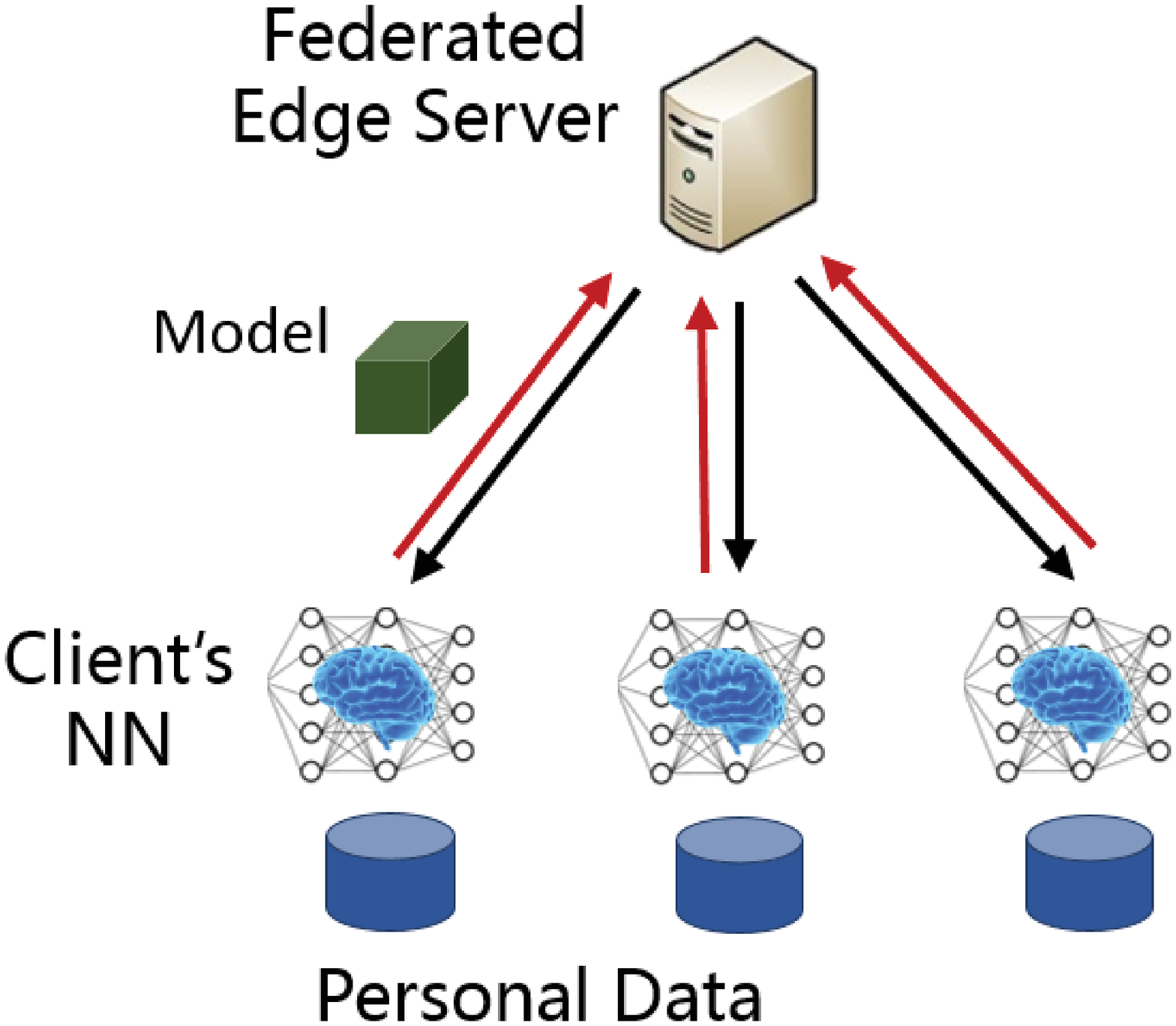} \label{aiplot5}}
\subfigure[]{\includegraphics[width=2in]{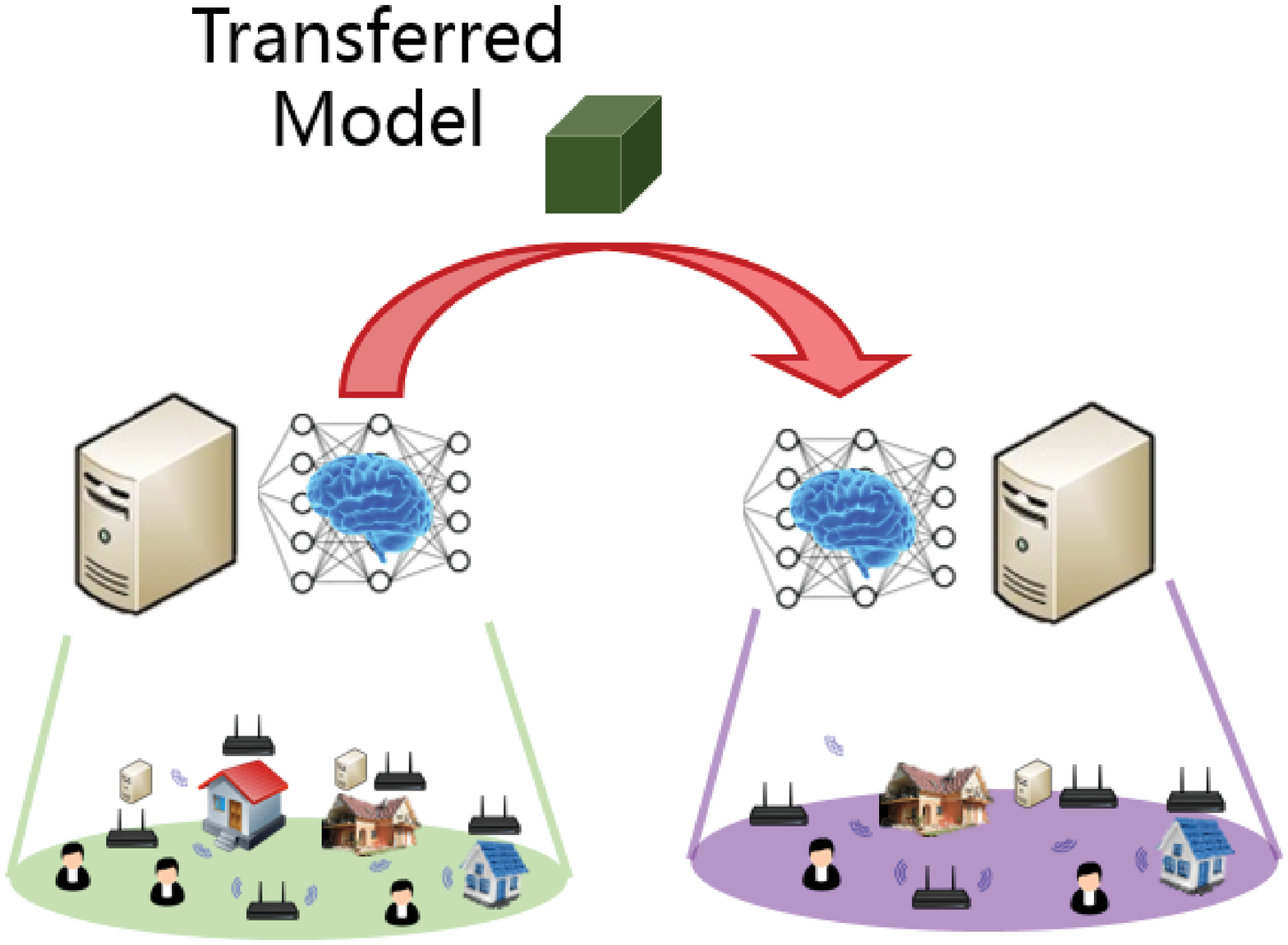} \label{aiplot6}}
\caption{Deep learning mechanisms of (a) supervisied, (b) unsupervised, and (c) reinforcement learning. The architecture of network learning of (d) distributed, (e) federated and (f) transfer learning.} \label{aisim}
\end{figure*}	

\section{Facet 5: Applications of Deep Learning in 6G Networks}
	Given the rapid development of AI-empowered deep learning, both \textit{supervised learning} as well as \textit{unsupervised learning} and \textit{reinforcement learning} have found favour in solving challenging communications and networking problems \cite{86,87}, as shown in Fig. \ref{aisim}. Specific examples are constituted by radio interference management, resource allocation, multiple parameter optimization \cite{88}, network traffic prediction, computing resource assignment, and flexible configuration of network functions \cite{ai}. In supervised learning, ground truth labels and fixed-size inputs constitute a deep layered neural network (NN). However, labeling is not required in unsupervised learning, which exploits the correlation between samples of the dataset. In reinforcement learning, an agent will interact with the environment and then updates the model based on the corresponding rewards. Note that deep learning can deal with comparably complex problems in a non-linear and non-convex manner than that utilizing machine learning. Therefore, we can efficiently manage both vertical and horizontal networks with the aid of deep neural networks. The AI schemes adopting deep learning for potential 6G applications and solutions are summarized at a glance in Table \ref{aitable}.
\begin{itemize}
\item \textbf{Supervised learning}: The open challenges in this technique include data collection and the appropriate data analytics in a practical network scenarios. Take wireless transmission for example, the collected beamforming data from real measurement or from solutions of convex optimization is served as ground-truth labels in an NN-based training. Moreover, laborious indoor fingerprinting for collecting signal features should be conducted for positioning and detection sensing. Several machine and deep learning techniques are designed to address different types of issues, such as 1) spatial correlation over data classification and prediction is well tackled by SVM, KNN, DNN and CNN, etc \cite{ourcsi1, ourcsi2, DNN1, 22-1}. 2) Temporal-domain problem is addressed by RNN and LSTM \cite{RNN1, RNN2, RNN3}. 3) Large-scale networking policy and management is perfectly performed by GNN \cite{GNN1, GNN2, GNN3}.

\item \textbf{Unsupervised learning}: The research target of this learning mechanism is focused on inference from an unlabelled dataset. Unlike supervised method with ground truth, unsupervised learning leverages iterative inference to attain hidden features for either dataset partitioning, clustering or augmentation. For example, GAN is promisingly adopted in network data augmentation \cite{GAN1, GAN2, GAN3} and security \cite{GAN4, GAN5, GAN6} for compensating information insufficiency from practical measurement. The highly-complex and high-dimensional large-scale network data processing can be well dealt with by using PCA method. While, accessing under uncertain and stochastic environment can rely on HMM or EM algorithms for maximizing total utility \cite{EM1}. The grand challenge lies on the accuracy and confidence of learning results, which becomes an open issue.

\item \textbf{Reinforcement learning}: Such technique is broadly employed for dynamic network policy adaptation in wireless network communication and computing resource management, network multi-parameters for diverse tele-traffic demands, QoS-guaranteed scheduling as well as accessing \cite{DQN1, DQN2, DQN3, DQN4}. Q learning uses a model-free mechanism by adjusting its policy according to updated system state and performance. However, its performance is limited by convergence speed and great uncertainties in a large-scale network. As an enhancement, DQN can tackle above-mentioned problems by using different model-based NNs for respective actions and evaluations based on interaction with the environment. However, the common challenge of reinforcement learning techniques lies on theoretical proofs of convergence, optimality and dynamic adaptation.

As a promising extension of DQN, \textit{deep deterministic policy gradient} (DDPG) based learning relies on a pair of neural networks forming an action-critic network: The action network provides the optimal policy, whilst the critic network evaluates the action. DDPG potentially enhances the stability, flexibility and adaptability to dynamic wireless communication systems \cite{ddpg1, ddpg3}. Furthermore, \textit{multi-agent reinforcement learning} \cite{marl1} is widely adopted in conjunction with multiple agents controlling their own policies, which mitigates the computational burden and memory requirements at the server. Note that multi-agent solutions may be viewed as multiple BSs and edges \cite{marledge}, a swarm of drones \cite{marluav} or vehicles \cite{marlcar}. They interact with the common shared environment and determine their next action without any information exchange overhead. To elaborate a little further, \textit{inverse reinforcement learning} (IRL), also referred to as learning from demonstrations \cite{irl0}, may also find applications in wireless communications and networking \cite{irl1}. In contrast to conventional forward reinforcement learning, IRL is capable of learning from an expert and may exhibit some human-like behaviors \cite{irl2}, which is popularly applied in robotic control systems.

%As a promising extension of DQN, \textit{deep deterministic policy gradient} (DDPG) enables two separated neural networks as action-critic networks: Action network provides the optimal policy, whilst critic network evaluates the action. DDPG potentially enhances the stability, flexibility and adaptability to dynamic wireless and communication systems \cite{ddpg1, ddpg3}. Furthermore, \textit{multi-agent reinforcement learning} \cite{marl1} is widely adopted with multiple agents controlling their own policies, which mitigates the computational burden and memory space at server. Note that multi-agent can be regarded as multiple BSs and edges \cite{marledge}, a swarm of drones \cite{marluav} or vehicles \cite{marlcar}. They act with the common shared environment and determine their next action without information exchange overhead. To elaborate a little further, \textit{inverse reinforcement learning} (IRL), so-called learning from demonstrations \cite{irl0}, is seeking its potentials in wireless communications and networking \cite{irl1}. Unlike conventional forward reinforcement learning, IRL is capable of learning from an expert, may have human-like behavior, to infer the corresponding reward function \cite{irl2}, which is commonly applied in robotic control. It is still under research how appropriately the wireless communications and networking experts can be defined in order to provide improved solutions. 
\end{itemize}
	
	Furthermore, the complexity of wireless propagation environments and the challenging requirements of high tele-traffic can be readily dealt with by sophisticated transfer learning methods \cite{TF1,TF2}. Briefly, \textit{transfer learning} directly employs the models that were previously trained under a network to a new environment for improving the efficiency of retraining \cite{89}. Given the rapid evolution of virtualized software-defined networks, it appears promising to adopt AI techniques for efficiently assigning resources both to central units as well as to edge servers and flexibly manage the resultant mobile network \cite{90}. In the past, fully-centralized computing was the norm, which often led to overloaded situations. As a remedy, \textit{distributed learning} allows the server to distribute its tasks to different computing units for parallel processing \cite{91}. For example, the 6G core network is expected to employ intelligent units for separately managing its control and user planes as well as the dynamic configuration of network functions \cite{92}.

	Some other new machine learning and deep learning techniques found in the AI-domain and applied for control and computer vision may also be harnessed for solving wireless communications and networking problems. As another attractive technique, \textit{meta learning} relies on sparse samples and labels for solving heterogeneous tasks \cite{meta0}. It can promptly adapt the trained parameters based on just a few experiences in a new environment at an impressive convergence rate and low computational complexity, which has already been employed in cellular \cite{meta1}, IoT \cite{meta2} and vehicular networks \cite{meta3}. AI techniques are also capable of efficiently processing massive amounts of IoT data \cite{93}, whilst the family of unlicensed access technologies may adopt AI-based learning models for detecting existing networks in order to avoid interference and packet collisions, while improving the spectrum versus energy efficiency \cite{94}. Moreover, deep neural networks are also capable of maintaining the QoS, while supporting numerous IoT devices \cite{95}. They can also preserve energy and adaptively collect environmental data \cite{96}. Additionally, we can also support network deployment as well as prediction and performance evaluation of a massive number of sensor nodes by employing graph convolutional networks \cite{97}. For ensuring network information security, \textit{federated learning} can be employed \cite{98,99}, where the encrypted models of local networks are uploaded and updated by a global controller, which prevents tapping and inference from the models by eavesdroppers. In the face of uncertainty, it is promising to conceive deep learning designs for ascertaining the authenticity of subscribers and for detecting unusual network behaviors, combat attacks from external networks, and ensure data privacy, especially, when relying on both limited computing resources and information \cite{100,101}.

\section{Summary and The Road to Multi-Component Pareto-Optimization}
	This article has listed five key research topics of next-generation wireless, including \textit{Next-Generation Architecture, Spectrum and Services}, \textit{Next-Generation Networking}, \textit{Internet of Things}, \textit{Wireless Positioning and Sensing}, as well as the \textit{Applications of Deep Learning in 6G Networks}. We have investigated comprehensive literature surveys for the potential promising techniques from the perspectives of architectures, networking, applications as well as scheme designs, which are extended from the current foundation of wireless and networking. We return to Fig. \ref{arc} and note that there is a potentially infinite number of system configurations that may be harnessed by next-generation systems. Throughout the past five generations there has been a gradual paradigm shift from bandwidth-efficiency relying on complex, high-delay near-capacity transceivers \cite{HanzoNearCap} towards maximizing the power-efficiency \cite{HanzoAchEE}, which has the fond connotation of green radio \cite{HanzoGreenRa}. With the introduction of 5G new radio and its URLLC mode, the importance of simultaneously maintaining both low latency and low bit error rate has reached the lime-light. This trend heralds a new multi-component optimization era \cite{HanzoSurveyMO}, in which the research community is expected to find all the so-called Pareto-optimal operating points of the associated multi-component objective functions, as highlighted for example in \cite{86, HanzoTwin, HanzoQuanMO, HanzoDeepMO}. 

\begin{figure}
	\centering
	\includegraphics[width=4.2in]{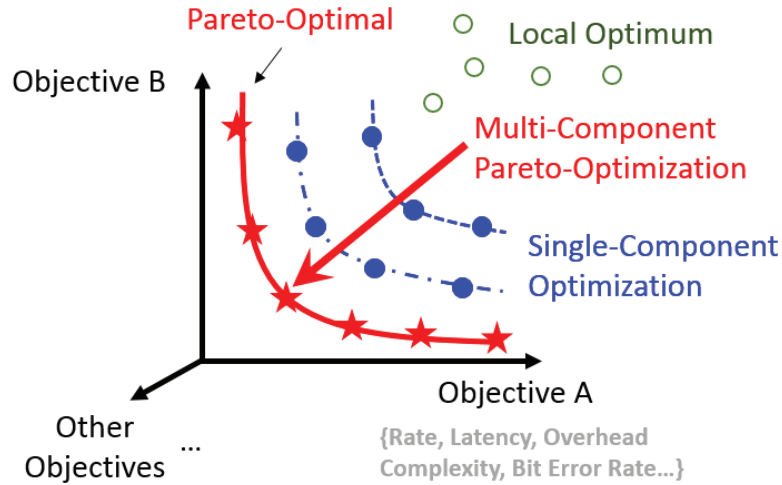}
	\caption{Illustration of a multi-component Pareto-optimization considering two objectives, such as rate, latency, overhead, complexity, and bit error rate metrics.} 
	\label{pare}
\end{figure}

	In this context the question arises: How can we boldly differentiate Pareto optimization from the set of simple conventional trade-offs? Explicitly, when we carry out for example single-component bandwidth-efficiency optimization, we completely ignore any other parameters or metrics of the system, such as its complexity or delay. In this context, we can for example always approach the Shannonian capacity more closely, if we employ a longer coding, which typically imposes an increased delay and escalating computing complexity. By contrast, in addition to this unconstrained Shannonian solution the Pareto front of all optimal configurations will contain the specific delay and complexity associated with each individual legitimate channel coded length. Hence, from physical layer perspective, this Pareto-optimal approach also goes way beyond the concept of finite-block-length information theory \cite{CodeFinite}, which simply quantifies the achievable performance associated with a specific coding-length, i.e., delay, but it remains oblivious of the complexity of a block-code capable of achieving it.
	
	To elaborate a little further in the context of a tangible example, let us assume that the multi-component objective function relying on bit error rate, throughput, delay as well as power has to be optimized. It is plausible that the throughput may be readily increased upon increasing the number of bits/symbol even without degrading the bit error rate, if we increase the power, i.e., degrade the power-efficiency and vice versa. Indeed, the throughput may also be improved without degrading the power-efficiency to the detriment of the bit error rate owing to increasing the number of bits/symbol. In this tangible practical context, the Pareto-front contains all optimal solutions. However, by definition none of the above parameters may be improved without degrading at least one of the others. Some other numerous tangible practical solutions may be found in \cite{HanzoNearCap, HanzoAchEE, HanzoGreenRa, HanzoSurveyMO, 86, HanzoTwin, HanzoQuanMO, HanzoDeepMO}.

	By relying on an asymptotic concept of \textit{multi-task learning} \cite{mtl1}, we are capable of incorporating powerful machine learning and deep learning techniques into multi-component optimization \cite{mtl2}. Depending on the specific requirements, multiple weighted loss functions can be designed for maximizing the detection accuracy and rate, while minimizing the processing time and energy in conjunction with four respective weights. AI-based multi-component optimization is capable of decomposing complex objective function spaces, striking a compelling tradeoff between processing efficiency and computational complexity \cite{mtl2}. As for applications in wireless communications and networking, federated learning and reinforcement learning based schemes are widely employed for multi-component learning optimization of diverse requirements, such as rate, throughput, latency, energy-spectrum efficiency and reliability \cite{mtl4,mtl5,mtl6,mtl7,mtl8}. Furthermore, transfer learning may be capable of resolving the dynamic multi-objective optimization problems routinely found in parameter initialization by the exploiting neural network's memory for faster convergence \cite{mtl3}. To conclude, a subset of open research issues in next-generation wireless are listed for providing insights gleaned from different fields, as the community moves from single-component to multi-component optimization. This radical system optimization principle may be expected to pervade the next-generation era, but requires a concerted community effort to make it a reality!

\bibliographystyle{IEEEtran}

%\footnotesize
\bibliography{myref}

\end{document}